\documentclass[journal,final,twocolumn,10pt,twoside]{IEEEtranTCOM}
\normalsize

\ifCLASSINFOpdf

\else
 
\fi

\usepackage{mathtools}
\usepackage{amsmath}
\usepackage{amssymb}
\usepackage{amsmath}
\usepackage{cite}
\usepackage{multirow}
\usepackage{booktabs}
\usepackage{array}
\usepackage{siunitx}
\usepackage{graphicx}
\usepackage{graphicx}

\date{}
\usepackage{mdwmath}
\usepackage{blindtext}
\usepackage{eqparbox}
\usepackage{fixltx2e}
\usepackage{color}
\usepackage{xcolor}
\usepackage{mdwmath}
\DeclareMathOperator{\E}{\mathbb{E}}
\usepackage{layout}
\begin{document}
\title{Performance Characterization of Relay-Assisted Wireless Optical CDMA Networks in Turbulent Underwater Channel}
\vspace{0ex}
\author{Mohammad Vahid Jamali, Farhad Akhoundi, and Jawad A. Salehi,~\IEEEmembership{Fellow,~IEEE} \vspace{0ex} 
\thanks{Part of this paper is
supported by Iran National Science Foundation (INSF). The authors are with the Optical Networks Research Laboratory (ONRL), Department of Electrical Engineering, Sharif University of Technology, Tehran, Iran (e-mail: mohammad.v.jamali@gmail.com; akhoundi.farhad@gmail.com ; jasalehi@sharif.edu).}
}

\maketitle
\begin{abstract}
\boldmath
In this paper, we characterize the performance of relay-assisted underwater wireless optical code division multiple access (OCDMA) networks over turbulent channels. In addition to scattering and absorption effects of underwater channels, we also consider optical turbulence as a log-normal fading coefficient in our analysis. To simultaneously and asynchronously share medium among many users, we assign a unique optical orthogonal code (OOC) to each user in order to actualize OCDMA-based underwater network. The most significant challenge in underwater optical communication is in the ability to extend the short range of its coverage. In order to expand the viable communication range, we consider multi-hop transmission to the destination. Moreover, we evaluate the performance of a relay-assisted point-to-point UWOC system as a special case of the proposed relay-assisted OCDMA network. Our numerical results indicate significant performance improvement by employing intermediate relays, e.g., one can achieve $32$ {dB} improvement in the bit error rate (BER) of $10^{-6}$ using only a dual-hop transmission in a $90$ {m} point-to-point clear ocean link.
\end{abstract}
\begin{IEEEkeywords}
 OCDMA, serial relaying, chip detect-and-forward, log-normal fading, BER performance, turbulent underwater channel, underwater wireless optical communications.
\end{IEEEkeywords}

\IEEEpeerreviewmaketitle

\section{Introduction}
\IEEEPARstart{U}{nderwater} wireless optical communication (UWOC) systems are receiving growing attention for various underwater applications. As opposed to their traditional counterparts, i.e., acoustic communications, they have three main superiorities: higher bandwidth, lower latency and better security. Therefore, UWOC system can be regarded as an alternative to meet the requirements of high speed and large data underwater communications such as imaging, real-time video transmission, high throughput sensor networks, etc \cite{tang2014impulse}. However, presently, UWOC is suitable only for ranges that are typically less than $100$ {m} which hinders its extensive usage. This drawback is due to the fact that UWOC suffers from three main impairing effects: absorption, scattering and turbulence which cause loss, inter-symbol interference (ISI) and fading on the received optical signal, respectively.

Many studies have been focused on characterizing absorption and scattering effects of different water types, both theoretically and experimentally \cite{mobley1994light,petzold1972volume}. Based on the experimental results reported in \cite{mobley1994light,petzold1972volume}, Tang-Dong-Zhang simulated UWOC channel by means of Monte Carlo (MC) approach considering absorption and scattering effects \cite{tang2014impulse}. They also fitted a double gamma function (DGF) to this impulse response and evaluated BER numerically, however in the absence of turbulence effects. Recently, Akhoundi-Salehi-Tashakori \cite{akhoundi2015cellular} proposed a cellular OCDMA-based UWOC network where a specified optical orthogonal code (OOC) {\cite{salehi1989code}} is assigned to each user. Furthermore, potential applications and challenges of such a network is elaborated in \cite{akhoundi2016cellular2016}.

 Although many valuable studies have been performed to characterize and mitigate turbulence-induced fading in free-space optical (FSO) communications \cite{navidpour2007ber,andrews2005laser,zhu2002free,karimi2009ber,karimi2011free}, its impairing effects on the performance of UWOC systems have received relatively less attention. However, some relevant and useful researches have recently been reported in the literature to characterize underwater fading statistics. For example in \cite{nikishov2000spectrum} an accurate power spectrum has been derived for fluctuations of turbulent seawater refractive index.
  Furthermore, Rytov method has been used in \cite{korotkova2012light} to evaluate the scintillation index of optical plane and spherical waves propagating in underwater turbulent medium. And in \cite{gerccekciouglu2014bit}, the on-axis scintillation index of a focused Gaussian beam has been formulated in weak oceanic turbulence and by considering log-normal distribution for intensity fluctuations the average BER in such systems is evaluated.

  It has been shown that the three aforementioned impairing factors in underwater medium are incremental functions of distance \cite{korotkova2012light,tang2014impulse}. This behaviour hinders on the system performance for longer ranges. To overcome this limitation and therefore to extend the viable communication range, in this paper we propose multi-hop transmission over turbulent underwater channel. By this scheme, we divide a relatively long communication distance with severe absorption, scattering and fading effects to shorter ones, each with a much reduced absorption, scattering and fading effects; and therefore with acceptable performance.
   Fig. 1 depicts an explanatory practical example of our proposed topology, namely underwater optical sensor network. Underwater sensor networks will acquire essential roles in investigating climate changes, disaster prevention, in monitoring biological, biogeochemical, evolutionary and ecological changes in the sea, ocean and lake environments, in pollution monitoring, and in helping to control and maintain oil production facilities. In this scheme, each sensor should ultimately upload its own data to the optical base transceiver station (OBTS)\footnote{The OBTS is located in the cell's center, and is connected to the other cells' OBTSs via fiber optic. Detailed experimental prototype of this underwater OBTS and the cellular network topology are presented in \cite{akhoundi2015cellular}}. Since these sensors are battery powered, their low-power signals will not withstand to reach a relatively far-located OBTS due to severe absorption, scattering and fading effects of the channel. In this circumstance, a comparatively simple relay can be implemented near this network to detect the received data and forward the detected data to the next relay (or to the OBTS in the last hop). Further, each sensor transmits its data based on OOC-OCDMA technique \cite{salehi1989code,salehi1989code2,kwong1991performance,prucnal2010optical,kwong2013optical} to introduce multiple access capability while reducing its interfering effect on the other sensors' signals. In general, in such difficult circumstances the relay-assisted strategy is a reasonable solution as opposed to designing an exclusive OBTS nearby the network, since relay nodes employment is much simpler and straightforward than OBTSs \cite{akhoundi2015cellular}. Moreover, different cells' OBTSs are sufficiently distant to provide longer range communications and also to reduce implementation costs. Hence, to develop temporary repeaters, implementation of simple relay nodes with narrow field of view (FOV) receivers and narrow/wide beam lasers/LEDs is more interesting than complicated omnidirectional OBTSs. Also the required processes in relays are much simpler than those of OBTSs, which involve different layers of network processes \cite{akhoundi2015cellular,akhoundi2016cellular2016}. In addition, our study of relay-assisted topology covers the more practical scenario of cooperative communication where different users operate as repeaters for each other.
           
 The rest of this paper is organized as follows. In Section II, the channel and system models in the context of relay-assisted OCDMA network are investigated. In Section III, some simplifying assumptions are made in order to approximate the BER of a relay-assisted underwater OCDMA network. Using Gaussian approximation to calculate the chip error rates (CERs) of different hops, the end-to-end BERs for both uplink and downlink transmissions are obtained. Further, we clarify how to obtain BER performance of a single-user relay-assisted point-to-point UWOC link as a special case of the aforementioned up-or downlink BERs. In Section IV, numerical results for various configurations are presented, and finally we conclude the paper in Section V.
\begin{figure}
  \centering
  \includegraphics[width=3.4in]{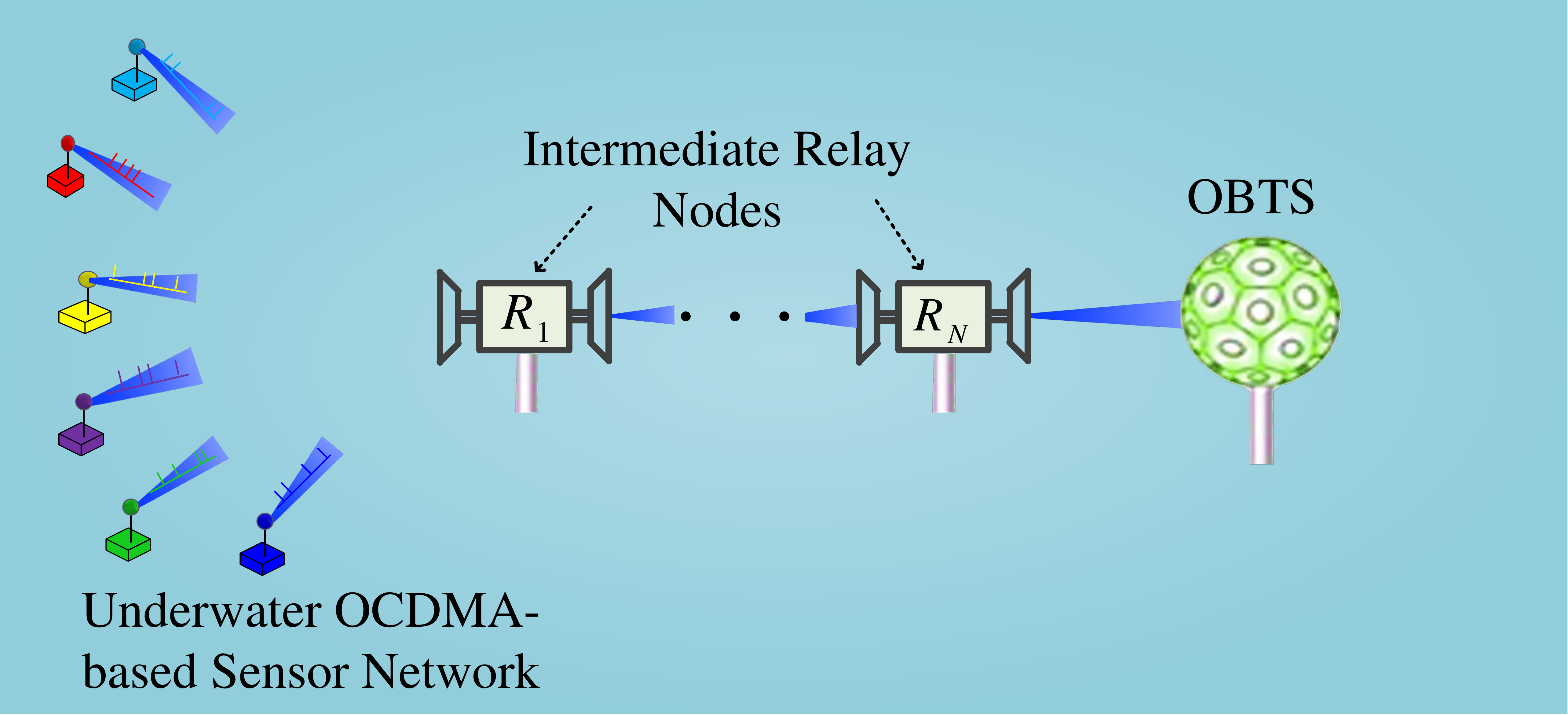}
  \caption{An illustrative practical example of relay-assisted OCDMA-based underwater wireless network: underwater optical sensor network.}
              \vspace{-0.15in}
  \end{figure}
\section{Channel and System Model}
\subsection{Channel Model}
 Through propagation of optical beam in underwater medium, interactions between each photon and seawater particles consist of absorption and scattering.
   Energy loss of non-scattered light, due to absorption and scattering can be characterized by absorption coefficient $a(\lambda )$ and scattering coefficient $b(\lambda )$, respectively. Further, extinction coefficient $c\left(\lambda \right)=a\left(\lambda \right)+b(\lambda )$ describes the total effect of absorption and scattering on energy loss. The values of these coefficients can vary with source wavelength $\lambda $ and water types \cite{tang2014impulse}. It has been shown in \cite{mobley1994light} that absorption and scattering have the lowest effect at the interval $400$ {nm} $<\lambda <530$ {nm}, so UWOC systems employ the blue/green region of the visible light spectrum to transmit and receive data.
In \cite{tang2014impulse} and \cite{cox2012simulation} the channel impulse response has been simulated based on MC approach with respect to the both absorption and scattering effects. In this paper the fading-free impulse response between any two $i$th and $j$th nodes is denoted by $h_{0,ij}(t)$. To attain this impulse response, we simulated the channel similar to \cite{tang2014impulse} and \cite{cox2012simulation} based on MC approach in order to thoroughly include absorption and scattering effects.

  In UWOC systems, optical turbulence is the main cause of fading on the propagating optical signal through turbulent seawater. Optical turbulence occurs due to random variations of refractive index. These random variations in underwater medium mainly results from fluctuations in temperature and salinity \cite{tang2013temporal}. To characterize turbulence effects, we multiply $h_{0,ij}\left(t\right)$ by a (positive) multiplicative fading coefficient ${\tilde{h}}_{ij}$ \cite{navidpour2007ber,andrews2005laser,zhu2002free}. Weak oceanic turbulence can be modeled with log-normal distribution \cite{gerccekciouglu2014bit,yi2015underwater} as;
 \begin{align} \label{pdf lognormal}
\!\!\!f_{{\tilde{h}}_{ij}}\!({\tilde{h}}_{ij} )=\frac{1}{2{{\tilde{h}}_{ij}}\sqrt{2\pi {\sigma }^2_{X_{ij}}}}{\rm exp}\!\left(\!\!-\frac{{\left({{\rm ln}  ({{\tilde{h}}_{ij}})\ }\!\!\!-\!2{\mu }_{X_{ij}}\right)}^2}{8{\sigma }^2_{X_{ij}}}\right)\!\!,
 \end{align}
 where ${\mu }_{X_{ij}}$ and ${\sigma }^{2}_{X_{ij}}$ are respectively the mean and variance of the Gaussian distributed log-amplitude factor $X_{ij}=\frac{1}{2}{\rm ln}({{\tilde{h}}_{ij}})$.
 To ensure that the fading coefficient conserves energy, we normalize fading amplitude such that $\E[{{{\tilde{h}}_{ij}} }]=1$, which implies ${\mu }_{X_{ij}}=-{\sigma}^2_{X_{ij}}$ \cite{navidpour2007ber}.
Therefore, in describing the fading statistics, it demands finding the dependency of log-amplitude variance ${\sigma}^2_{X_{ij}}$ to the ocean turbulence parameters where we will discuss in the remaining of this subsection.

 The scintillation index of a light wave with intensity $I_{ij}=I_{0,ij}{{\tilde{h}}_{ij}}$ is defined by \cite{korotkova2012light,andrews2005laser};
 \begin{align} \label{S.I.}
 {\sigma^2_{I_{ij}}}=\frac{\E[{I^2_{ij}}]-\E^2[{I_{ij}}]}{\E^2[{I_{ij}}]}=\frac{\E[{{\tilde{h}}^2_{ij}}]-\E^2[{{\tilde{h}}_{ij}}]}{\E^2[{{\tilde{h}}_{ij}}]},
 \end{align}
in which $I_{0,ij}$ is the fading-free intensity.
{Scintillation index can be considered as a threshold for separating the weak and strong turbulence regimes, i.e., when this parameter is smaller than unity the channel is in weak turbulence and vice versa \cite{andrews2005laser,yi2015underwater,korotkova2012light}.}
 It has been shown in \cite{korotkova2012light} that under the assumption of weak turbulence the scintillation index of plane and spherical waves can be calculated as;
\begin{align}
 &\sigma^2_I=8{\pi}^2{k_0}^2d_0\int_{0}^{1}\int_{0}^{\infty}\kappa{\Phi}_n(\kappa)\bigg[1-\nonumber\\
 &~~~~~~~~~~~~{\rm cos}\left(\frac{d_0{\kappa}^2}{k_0}\xi\left(1-\left(1-\Theta\right)\xi\right)\right)  \bigg]d\kappa d\xi,
\end{align}
\noindent where $\Theta=1$ and $0$ for plane and spherical waves, respectively. $k_0={2\pi }/{\lambda }$, $d_0$ and $\kappa$ denote the wave number, propagation distance and scalar spatial frequency, respectively. Moreover, ${\Phi }_n(\kappa )$ is the power spectrum of turbulent fluctuations which has the form \cite{nikishov2000spectrum,korotkova2012light};
\begin{align} \label{phi_n}
{\Phi }_n\left(\kappa \right)=& 0.388\times {10}^{-8}{\varepsilon }^{-{1}/{3}}{\kappa }^{{-11}/{3}}\left[1+2.35{\left(\kappa \eta \right)}^{{2}/{3}}\right]\nonumber\\
&\times\frac{{\chi }_T}{w^2}\left(w^2e^{-A_T\delta }+e^{-A_S\delta }-2we^{-A_{TS}\delta }\right),
\end{align}
\noindent where $\varepsilon $ is the rate of dissipation of turbulent kinetic energy per unit mass of fluid, $\eta ={{10}^{-3}}$ {m} is the Kolmogorov micro-scale, ${\chi }_T$ is the rate of dissipation of mean-square temperature and $w$ is the relative strength of temperature and salinity fluctuations which takes on values in the interval $\left[-5,0\right]$.
Other parameters in \eqref{phi_n} are as $A_T=1.863\times {10}^{-2}$, $A_S=1.9\times {10}^{-4}$, $A_{TS}=9.41\times {10}^{-3}$ and $\delta =8.284{\left(\kappa \eta \right)}^{{4}/{3}}+12.978{\left(\kappa \eta \right)}^2$ \cite{nikishov2000spectrum,korotkova2012light}.
 It can be shown that for turbulent channels with log-normal fading distribution the scintillation index is related to the log-amplitude variance as $\sigma^2_{I_{ij}}={\exp}(4{\sigma}^2_{X_{ij}})-1$ \cite{andrews2005laser}. Therefore, for relatively weak oceanic turbulence the scintillation index (which is comprehensively evaluated in \cite{korotkova2012light}) directly yields the log-amplitude variance and consequently specifies fading coefficients' probability density function (PDF) in \eqref{pdf lognormal}.

{It is worth noting that given a set of conditions, e.g., the link range, the water type, inherent optical properties (IOPs), etc., each of the aforementioned three impairing effects of the channel can be a serious concern. For example, as the water turbidity increases the absorption and scattering coefficients increase too. Therefore, in turbid harbor waters absorption and scattering are very important and should be treated, e.g., using multi-hop transmission. On the other hand, the severity of turbulence, which is measured by the scintillation index, depends on the values of the parameters $\chi_T$, $\varepsilon$ and $w$. Based on the experimental results in \cite{thorpe2007introduction} the typical values of $\varepsilon$ are between $10^{-8}$ ${\rm m^2/s^3}$ to $10^{-2}$ ${\rm m^2/s^3}$. Also $\chi_T$ takes the values between $7\times10^{-10}$ ${\rm K^2/s}$ to $10^{-4}$ ${\rm K^2/s}$. Hence, depending on the values of these parameters, strong turbulence can occur at ranges as short as $10$ {m} and as long as $100$ {m} \cite{korotkova2012light}, which impressively differs from atmospheric channels where strong turbulence distances are on the order of kilometers. Therefore, alleviating such a strong turbulence requires more studies.}
\subsection{System Model}
\begin{figure}
     \centering
     \includegraphics[width=3.4in]{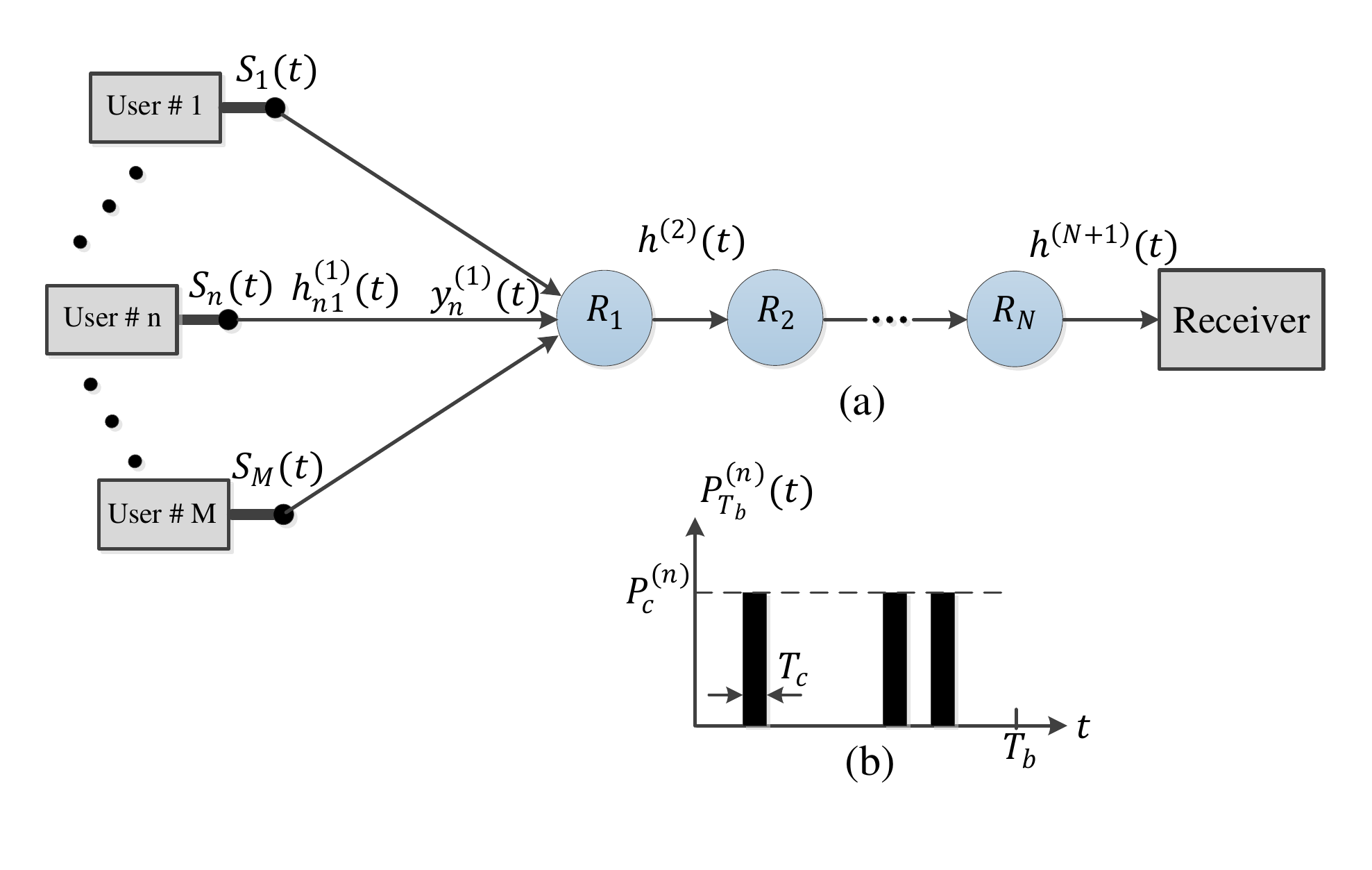}
     \caption{(a) Block diagram of the relay-assisted OCDMA-based underwater wireless network; (b) transmitted signal of the $n$th user in its $0$th time slot.}
                 \vspace{-0.15in}
     \end{figure}
We consider a collection of $M$ underwater users which are located nearby each other and are communicating with a relatively distant receiver or OBTS. To simultaneously and asynchronously share the medium among the various users, a specific OOC code with weight $W$ and length $F=T_b/{T_c}$ is assigned to each user, where $T_b$ and $T_c$ are bit and chip duration times, respectively. To guarantee that auto-and crosscorrelation constraints $\lambda_a$ and $\lambda_c$ are equal to unity, maximum allowable number of users is upper bounded by;
\begin{align} \label{M user}
M\leq \left \lfloor \frac{F-1}{W{(W-1)}} \right \rfloor,
\end{align}
where $ \left \lfloor x \right \rfloor$ is the integer portion of the real value $x$ \cite{salehi1989code}.

 As stated in the previous section absorption, scattering and fading effects are distance-dependent and their severity increases for longer ranges. This issue motivates the idea of placing intermediate nodes, namely relay nodes, between the users and the destination to alleviate the above mentioned disturbing effects. Therefore, as illustrated in Fig. 2(a) we have located $N$ intermediate relays in a serial topology between the users and the destination to increase the viable communication range. This topology is also suitable when there is no line-of-sight to the destination. In order to simplify relays' structure, chip detect-and-forward (CDF) strategy is employed at the relay nodes. In this scheme, each relay simply detects the received chips based on an appropriate threshold and forwards the detected chips to the next (or final) node. It is obvious that chip detection process in the first relay $R_1$ is affected by multiple access interference (MAI); however, this interference {does not} affect the chip detection process in the remaining relays and the destination, since each of these nodes only receives the transmitted signal by the previous node.

As it is shown in Fig. 2(b), the $n$th user's $0$th time slot transmitted OCDMA-based pulse shape, $P_{T_b}^{(n)}(t)$, has the form;
\begin{align} \label{P_T^n(t)}
P_{T_b}^{(n)}(t)=P_c^{(n)}\sum_{j=0}^{F-1}C_j^{(n)}P_{T_c}(t-jT_c),
\end{align}
where $P_c^{(n)}$ is the $n$th user's chip power and is related to the average transmitted power per bit of this user as $P_{ave}^{(n)}=\frac{W}{2F}P_c^{(n)}$. And $C^{(n)}=(C_0^{(n)},C_1^{(n)},...,C_{F-1}^{(n)})$ is the $n$th user's OOC code with length $F$ and weight $W$ \cite{salehi1989code2}. $P_{T_c}(t)$ is a rectangular pulse with duration $T_c$ and unit amplitude, i.e., $P_{T_c}(t)=\Pi(\frac{t-{(T_c/2)}}{T_c})$, where $\Pi(t)$ is a rectangular pulse with unit amplitude in the interval $[-1/2,1/2]$. Therefore, the transmitted OCDMA-based signal of the $n$th user, $S_n(t)$, can be expressed as;
\begin{align} \label{S_n(t)}
 S_n(t)&=\sum_{k=-\infty}^{\infty}b_k^{(n)}P_{T_b}^{(n)}(t-kT_b)\nonumber\\
 &=P_c^{(n)}\sum_{k=-\infty}^{\infty}b_k^{(n)}\sum_{j=0}^{F-1}C_j^{(n)}P_{T_c}(t-kT_b-jT_c),
\end{align}
 in which $b_k^{(n)}\in\left \{ 0,1 \right \}$ is the $n$th user's transmitted bit on its $k$th time slot, representing on-off keying (OOK) modulation. Furthermore, the received optical signal from the $n$th user to $R_1$, after having propagated through the channel with impulse response of ${h}^{(1)}_{n1}(t)=\tilde{h}^{(1)}_{n1}{h}^{(1)}_{0,n1}(t)$, can be represented as;
 \begin{align} \label{y_n^{(1)}(t)}
& y_n^{(1)}(t)=S_n(t){\ast}{h}^{(1)}_{n1}(t)\nonumber\\
& ~~~=P_c^{(n)}\tilde{h}^{(1)}_{n1}\!\!\!\sum_{k=-\infty}^{\infty}\!\!\!b_k^{(n)}\!\sum_{j=0}^{F-1}\!C_j^{(n)}\Gamma^{(1)}_{n,1}(t\!-\!kT_b\!-\!jT_c\!-\tau_n),
 \end{align}
 where ${h}^{(1)}_{n1}(t)$ is the aggregated channel impulse response of the first hop from the $n$th user to the first relay, $\Gamma^{(1)}_{n,1}(t)=P_{T_c}(t)\ast{h}^{(1)}_{0,n1}(t)$, and $\ast$ denotes the convolution operation. $\tau_n$ is the $n$th user's delay time with respect to the first relay receiver's reference clock. In this paper we consider $\tau_n$ to be chip synchronous with an integer multiple of $T_c$ \cite{salehi1989code2}. Note that each of the $M$ users produces similar contribution in the first relay. Hence, the total received optical signal at the front end of the first relay's detector can be summarized as;
 \begin{align} \label{y^1(t)}
  y^{(1)}(t)=\sum_{n=1}^{M}y_n^{(1)}(t).
  \end{align}
  Moreover, various types of noises, i.e., background light, dark current and thermal noise all affect the system performance. In this paper we adopt Poisson distributed background light and dark current and Gaussian distributed thermal noise \cite{einarsson2008principles}.

{We should emphasize that our system model and analytical derivations are general and valid for when the transmitters use collimated laser beams or diffusive light emitting diodes (LEDs). However, from the practical point of view, since the users can be mobile, tracking and pointing issues suggest to use diffusive links for the first hop. In other words, in order to reduce the pointing and tracking difficulties, it is preferable to use LEDs or lasers with enough wide divergence angles as transmitters of the first hop, during both up-and downlink transmissions. Also due to the same reason, receivers with larger FOV are preferable for the first hop. On the other hand, since relay nodes are fixed, the pointing and tracking troubles decrease for the intermediate hops. Therefore, collimated links using narrow beam divergence lasers are more suitable for these hops to better save energy and reduce the channel temporal spreading. Further, if the destination node is a fixed simple receiver, collimated link is more favorable to better save energy and mitigate ISI. Finally, for the depicted topology in Fig. 1, where destination is an OBTS, due to the physical properties of OBTS \cite{akhoundi2015cellular} the transmitter of the last hop must have relatively a wide beam divergence angle to better illuminate the separate photodetectors of the OBTS.}
 \section{BER Performance Analysis}
 To evaluate BER, we assume the first user as the desired user and the remaining users as undesired interfering users. Hence, the average BER of the desired user under the assumption that all users transmit bits ``$0$" and ``$1$" with equal probability, can be evaluated as;
 \begin{align} \label{P_E}
 P_E=\frac{1}{2}P_{be}(1|0)+\frac{1}{2}P_{be}(0|1),
 \end{align}
 in which $P_{be}(1|0)$ and $P_{be}(0|1)$ are bit error probabilities when bits ``$0$" and ``$1$" are transmitted, respectively. To obtain $P_{be}(1|0)$ and $P_{be}(0|1)$, we first make the following two assumptions:
 \subsubsection{No Inter-Symbol Interference (ISI)}
We assume that all of the users transmit with a chip rate in which the main portion of the energy of the received optical signal $\Gamma(t)=P_{T_c}(t)\ast h_0(t)$ is confined in the interval $(0,T_c)$, i.e.,
\begin{align} \label{compare}
\int_{0}^{T_c}\Gamma(t)dt\gg\int_{-\infty}^{0}\Gamma(t)dt+\int_{T_c}^{\infty}\Gamma(t)dt.
\end{align}
 Under this assumption we can conclude that ISI has a negligible effect on the system performance. Further, in order to more accurately take into account the loss due to absorption and scattering, we simulate the channel impulse response by MC approach and we guarantee \eqref{compare} by adopting users' data rate not to exceed the predetermined value indicated by \eqref{compare}. Taking into account that the channel delay spread is much smaller than $T_c$, the channel loss coefficient on a given chip can be determined as;
 \begin{align} \label{loss}
 L=\frac{\int_{0}^{T_c}\Gamma(t)dt}{\int_{0}^{T_c}P_{T_c}(t)dt}=\frac{1}{T_c}\int_{0}^{T_c}\Gamma(t)dt.
 \end{align}
 
 {It is worth mentioning that as the link range, water turbidity and transmitter beam divergence angle increase the channel imposes more attenuation (loss) and time spreading on the propagating optical signal. On the other hand, as the receiver FOV increases the channel spreading time increases too. In these conditions, the assumption in \eqref{compare} leads to larger values for $T_c$ and hence limits the transmission data rate to smaller values.}
 \subsubsection{Identical Chip Power}
 We assume that all users have the same transmit power as $P_c$. Furthermore, each relay after detecting the received chips, forwards each ``ON" chip with power $P_c$.
 
 In what follows, we apply the aforementioned two assumptions to analyze the BERs of the uplink and downlink transmissions.
 \subsection{Uplink BER Analysis}
  In this section we focus on the $0$th time slot transmitted bit of the first (desired) user. All the other users transmit their {signals} asynchronously which cause MAI on the desired user. Therefore, with respect to the aforementioned two assumptions and \eqref{y^1(t)}, we can express the uplink received signal at $R_1$ as;
  \begin{align} \label{y^{(1)}(t)-2}
   & y^{(1)}(t)= P_c\tilde{h}^{(1)}_{11}b_0^{(1)}\sum_{j=0}^{F-1}C_j^{(1)}\Gamma_{1,1}^{(1)}(t-jT_c-\tau_1)\nonumber\\
   & \!+P_c\!\sum_{n=2}^{M}\tilde{h}^{(1)}_{n1}\!\!\!\sum_{k=-\infty}^{\infty}\!\!\!b_k^{(n)}\!\sum_{j=0}^{F-1}\!C_j^{(n)}\Gamma_{n,1}^{(1)}(t\!-\!kT_b\!-\!jT_c\!-\!\tau_n),\!\!
   \end{align}
  which is composed of two main parts: desired and the undesired MAI signals. Let us define $\vec{\alpha}=(\alpha_1,\alpha_2,...,\alpha_W)$ as the interference
  pattern occurred on the pulsed mark chips of the first user's OOC, where $\alpha_q$ is the number of interferences on the $q$th pulsed mark chip of the desired user's OOC. Furthermore, we define $l$ as the total number of interferences that occurred on all the chips of the desired user, i.e., $l=\sum_{q=1}^{W}\alpha_q$. The random variable (RV) $l$, which specifies the number of interfering users on the desired user, has a Binomial distribution function. The joint probability distribution function of the interference pattern can be represented as \cite{ghaffari2008wireless,zahedi2000analytical};
  \begin{align} \label{Pr(alpha,l)}
  P_l(l,\vec{\alpha})=\frac{l!}{W^l\prod_{q=1}^{W}(\alpha_q)!}\!\binom{M-1}{l}\!\!\left(\frac{W^2}{2F}\right)^l\!\!\left(\!1\!-\!\frac{W^2}{2F}\right)^{M-1-l}\!\!.
  \end{align}
  Therefore to evaluate \eqref{P_E}, or equivalently $P_{be}(\overline{b_0^{(1)}}|b_0^{(1)})$, we need to evaluate the BER conditioned on $l$ and $\vec{\alpha}$. Then, the unconditioned error probability can be evaluated as;
  \begin{align} \label{unconditioned}
 P_{be}(\overline{b_0^{(1)}}|b_0^{(1)})=\sum_{l=0}^{M-1}\sum_{\vec{\alpha}\in\Omega_l}P_{be}(\overline{b_0^{(1)}}|b_0^{(1)},l,\vec{\alpha})P_l(l,\vec{\alpha}),
  \end{align}
  where $\overline{b_0^{(1)}}$ is the binary complement of the $0$th time slot transmitted bit of the first user ${b_0^{(1)}}$ and $\Omega_l$ is the set of all interference patterns with $l$ interfering users. Moreover, to evaluate $P_{be}(\overline{b_0^{(1)}}|b_0^{(1)},l,\vec{\alpha})$ we first need to calculate conditioned on fading coefficients vector $\bar{H}$ and then average the conditional BER over $\bar{H}$ as;
  \begin{align} \label{averaging}
  \!\!P_{be}(\overline{b_0^{(1)}}|b_0^{(1)},l,\vec{\alpha})\!=\!\!\int_{\bar{H}}\!P_{be}(\overline{b_0^{(1)}}|b_0^{(1)},l,\vec{\alpha},\bar{H})f(\bar{H})d\bar{H},\!\!
  \end{align}
  where $f(\bar{H})$ is the joint PDF of fading coefficients in $\bar{H}$. Note that $\bar{H}$ may be different for each of up-and downlink transmissions. We will specify this vector at the end of each subsection for each of configurations.

{Different rules can be adopted at the receiver side to decide on the transmitted data bits regarding the detected chips. Due to the positive nature of incoherent OCDMA systems, the presence of interfering users manifests its effect as an increase on the photoelectrons count. Therefore, MAI is more likely to cause errors on the chips ``OFF" rather than on the chips ``ON". In order to mitigate this asymmetry in error probabilities for chips ``ON" and ``OFF", we employ the methodology introduced in \cite{zahedi2000analytical} and \cite{shalaby1998chip}. The methodology is based on the following decision rule which imposes more restrictions on the correct detection of bit $``1"$: decide that the transmitted bit was ``$1$" if all detected chips are ``ON" and otherwise recognize ``$0$" as the transmitted data bit. In this case, conditional error probabilities in \eqref{averaging} can be characterized as follows;}
  \begin{subequations} \label{P_be}
  \begin{align}
 & P_{be}(1|0,l,\vec{\alpha},\bar{H})=\prod_{q=1}^{W}P^{(q)}_{ce-e2e}(1|0,l,\vec{\alpha},\bar{H}),\\
 & P_{be}(0|1,l,\vec{\alpha},\bar{H})=\!1\!-\!\prod_{q=1}^{W}\!\left[1\!-\!P^{(q)}_{ce-e2e}(0|1,l,\vec{\alpha},\bar{H})\right],\!\!
  \end{align}
  \end{subequations}
in which $P^{(q)}_{ce-e2e}(1|0,l,\vec{\alpha},\bar{H})$ and $P^{(q)}_{ce-e2e}(0|1,l,\vec{\alpha},\bar{H})$ are conditional end-to-end chip error rates on the $q$th transmitted chip of the desired user when this chip is ``OFF" and ``ON", respectively. To obtain approximate end-to-end CERs, we make the following assumption: although it is possible to detect a chip correctly at the receiver despite of incorrect detection in some of the intermediate relays, we neglect these lucky events and assume that a chip can be detected correctly at the receiver if and only if it is transmitted without any error over all intermediate hops. Then, we have;
\begin{subequations} \label{p_{ce-e2e}}
   \begin{align}
  & \!\!\!P^{(q)}_{ce-e2e}(1|0,\!l,\!\vec{\alpha},\!{\bar{H}})\!=\!1\!-\!\!\!\prod_{i=1}^{N+1}\!\left[1\!-\!P^{(q)}_{ce-i}(1|0,l,\vec{\alpha},{\tilde{h}}^{(i)})\right]\!,\!\! \\
  & \!\!\!P^{(q)}_{ce-e2e}(0|1,\!l,\!\vec{\alpha},\!{\bar{H}})\!=\!1\!-\!\!\!\prod_{i=1}^{N+1}\!\left[1\!-\!P^{(q)}_{ce-i}(0|1,l,\vec{\alpha},{\tilde{h}}^{(i)})\right]\!,\!\!
   \end{align}
   \end{subequations}
  where ${\tilde{h}}^{(i)}$ is the $i$th hop fading coefficient\footnote{Note that the fading coefficient of the first hop in uplink transmission involves more than one coefficient (specifically $\tilde{h}^{(1)}_{11}$ and $\beta_q^{(I)}$), as will be demonstrated in Section III-A1.}. $P^{(q)}_{ce-i}(1|0,l,\vec{\alpha},{\tilde{h}}^{(i)})$ and $P^{(q)}_{ce-i}(0|1,l,\vec{\alpha},{\tilde{h}}^{(i)})$ are the conditional $q$th chip errors of the $i$th hop when chips ``OFF" and ``ON" are transmitted, respectively. Therefore, to determine \eqref{P_be} and \eqref{p_{ce-e2e}} we begin by calculating CER of each intermediate hop.
 \subsubsection{Conditional CER Analysis for the First Hop}
 Here we calculate the conditional chip error probabilities of the first hop using Gaussian approximation, which provides an acceptable estimate of the system error rate specially for error probability values less than $0.1$ \cite{einarsson2008principles}. The first relay's receiver simply integrates over each $T_c$ seconds and compares the result with an appropriate threshold to detect the received chip \cite{shalaby1998chip}. In addition to this signal, Poisson distributed dark current and background light (with mean photoelectron count rates of $\gamma_d$ and $\gamma_b$, respectively) and Gaussian distributed thermal noise are added to the received signal count in each $T_c$ seconds. Therefore, photo-detected signal after integrate-and-dump circuit of the first relay can be modeled as;
 \begin{align} \label{r_1}
\overrightarrow{r_{1}}=\overrightarrow{y_{1}}+\overrightarrow{v_{1}},
 \end{align}
 in which $\overrightarrow{v_{1}}$ is a vector with $W$ uncorrelated Gaussian distributed elements each with mean zero and variance of;
 \begin{align}\label{thermal}
 \sigma^2_{th}=\frac{2K_bT_rT_c}{Re^2},
 \end{align}
 corresponding to the integrated thermal noise \cite{jazayerifar2006atmospheric}.
 $K_b$, $e$, $T_r$, and $R$ are Boltzmann's constant, electron's charge, the receiver
 equivalent temperature, and load resistance, respectively.
 On the other hand, $\overrightarrow{y_{1}}$ is an array of Poisson distributed RVs with mean $\overrightarrow{m_{1}}$ as;
\begin{align} \label{m_1}
 \overrightarrow{m_{1}}={\overrightarrow{m_{1}}}^{(d)}+{\overrightarrow{m_{1}}}^{(I)}+{\overrightarrow{m_{1}}}^{(bd)},
 \end{align} 
  where based on \eqref{y^{(1)}(t)-2}, ${\overrightarrow{m_{1}}}^{(d)}$ corresponds to the desired count that it is obtained as;
    \begin{align} \label{m_1^d}
    {\overrightarrow{m_{1}}}^{(d)}=\frac{\eta}{hf}\tilde{h}^{(1)}_{11}b_0^{(1)}P_cL_{1,1}^{(1)}T_c\vec{u},
    \end{align}
    in which $\eta$, $h$, and $f$ are the photodetector's quantum efficiency, Planck's constant, and the optical frequency, respectively. $L_{1,1}^{(1)}$ is the aggregated channel loss (due to absorption and scattering effects) of the first hop from the desired user to the first relay, and $\vec{u}=(1,1,...,1)$ is a $W$-dimensional all-one vector.
Moreover, ${\overrightarrow{m_{1}}}^{(I)}$ that corresponds to the count of the undesired interfering users can be expressed as;
\begin{align} \label{m_1^I}
 {\overrightarrow{m_{1}}}^{(I)}=\frac{\eta}{hf}P_cT_c{\overrightarrow{\beta}}^{(I)},
 \end{align} 
 where ${\overrightarrow{\beta}}^{(I)}$ is a vector with length $W$ which its $q$th element $\beta_q^{(I)}$ is the weighted sum of $\alpha_q$ independent log-normal RVs, corresponding to the sum of interfering users' fading coefficients, i.e.,
 \begin{align} \label{sum_log}
 \beta_q^{(I)}=\sum_{n\in{\Lambda_q}}L_{n,1}^{(1)}\tilde{h}_{n1}^{(1)},
 \end{align}
 in which $L_{n,1}^{(1)}$ is the aggregated channel loss of the first hop from the $n$th user to the first relay. ${\Lambda_q}$ specifies the set of $\alpha_q$ interfering users on the $q$th chip. Weighted sum in \eqref{sum_log} can be approximated as a single log-normal RV $\beta_q^{(I)}\approx{\exp}(2z_q^{(I)})$ with log-amplitude mean $\mu_{z_q^{(I)}}$ and variance $\sigma^2_{z_q^{(I)}}$ as \cite{safari2008relay};
   \begin{align}
   &\mu_{z_q^{(I)}}=\frac{1}{2}{\rm ln}\bigg(\sum_{n\in{\Lambda_q}}L_{n,1}^{(1)}\bigg)-\sigma^2_{z_q^{(I)}}, \label{mu_z}\\
    & \sigma^2_{z_q^{(I)}}=\frac{1}{4}{\rm ln}\left(1+\frac{\sum_{n\in{\Lambda_q}}{\left(L_{n,1}^{(1)}\right)}^2\left(e^{4\sigma^2_{X_{n1}}}-1\right)}{{\left(\sum_{n\in{\Lambda_q}}L_{n,1}^{(1)}\right)}^2}\right) \label{sigma_z},
    \end{align}
 where $\sigma^2_{X_{n1}}$ is the variance of log-amplitude factor $X_{n1}^{(1)}=\frac{1}{2}{\rm ln}(\tilde{h}_{n1}^{(1)})$. Furthermore, for $\alpha_q=0$ there is not any interference on the $q$th pulsed mark chip and $\Lambda_q$ in \eqref{sum_log} specifies an empty set and consequently $\beta_q^{(I)}=0$. On the other hand, ${\overrightarrow{m_{1}}}^{(bd)}=(\gamma_d+\gamma_b)T_c\vec{u}$ in \eqref{m_1} corresponds to the dark current and background light counts vector.
 We can summarize the first user's uplink average received counts of the $q$th chip conditioned on transmitting chips ``OFF" and ``ON" respectively as;
 \begin{subequations} \label{23}
 \begin{align}
 & m_{0,q}^{(ul,1)}=\frac{\eta}{hf}P_cT_c\beta_q^{(I)}+(\gamma_d+\gamma_b)T_c,\\
  & m_{1,q}^{(ul,1)}=\frac{\eta}{hf}P_cT_c\left[L_{1,1}^{(1)}\tilde{h}^{(1)}_{11}+\beta_q^{(I)}\right]+(\gamma_d+\gamma_b)T_c.
 \end{align}
 \end{subequations}
 
{{In order to evaluate the first hop's conditional CER, we apply Gaussian approximation which approximates  all Poisson distributed RVs with Gaussian distributed RVs, where each of these Gaussian variables has equal mean and variance.
The receiver of the first hop selects the threshold based on Gaussian approximation as $\Theta_g={\left(\sigma_0E_1+\sigma_1E_0\right)}/{\left(\sigma_1+\sigma_0\right)}$,
  where $\sigma_0$ and $\sigma_1$ are standard deviations, and $E_0$ and $E_1$ are means of the Gaussian distributed decision variable when transmitted chips are ``OFF" and ``ON", respectively \cite{einarsson2008principles}. Note that the first hop's receiver is synchronized with the desired user and hence is imperceptive to the transmitted chips of interfering users. Also we assume that the channel state information (CSI) is available at the receiver, i.e., the first hop's receiver perfectly knows $\tilde{h}^{(1)}_{11}$. Therefore, based on Eqs. \eqref{r_1}-\eqref{23} and Gaussian approximation \cite{einarsson2008principles}, $E_0=(\gamma_d+\gamma_b)T_c$, $E_1=\frac{\eta}{hf}P_cT_cL_{1,1}^{(1)}\tilde{h}^{(1)}_{11}+E_0$, $\sigma_0=\sqrt{E_0+\sigma^2_{th}}$ and $\sigma_1=\sqrt{E_1+\sigma^2_{th}}$. Also $\Theta_g$ simplifies to $\Theta_g=\sqrt{(E_1+\sigma^2_{th})(E_0+\sigma^2_{th})}-\sigma^2_{th}$.}}
 
 {{Now we can obtain the first hop $q$th chip conditional error probabilities as;
 \begin{subequations} \label{p-ce-1-new}
  \begin{align}
  \!\!P^{(q,ul)}_{ce-1}(1|0,l,\vec{\alpha},\tilde{h}^{(1)}_{11},\beta_q^{(I)})\!&=\Pr(\boldsymbol{U^{(1)}_0}>\Theta_g|l,\vec{\alpha},\tilde{h}^{(1)}_{11},\beta_q^{(I)})\nonumber\\
  &=Q\left(\frac{\Theta_g-m_{0,q}^{(ul,1)}}{\sqrt{m_{0,q}^{(ul,1)}+\sigma^2_{th}}}\right),\!\!\\
  \!\!P^{(q,ul)}_{ce-1}(0|1,l,\vec{\alpha},\tilde{h}^{(1)}_{11},\beta_q^{(I)})\!&=\Pr(\boldsymbol{U^{(1)}_1}\leq\Theta_g|l,\vec{\alpha},\tilde{h}^{(1)}_{11},\beta_q^{(I)})\nonumber\\&=Q\left(\frac{m_{1,q}^{(ul,1)}-\Theta_g}{\sqrt{m_{1,q}^{(ul,1)}+\sigma^2_{th}}}\right),\!\!
  \end{align}
  \end{subequations}
  in which $\boldsymbol{U^{(1)}_0}$ and $\boldsymbol{U^{(1)}_1}$ are Gaussian distributed decision variables of the first receiver, when the transmitted chips are ``OFF" and ``ON", respectively. $\boldsymbol{U^{(1)}_0}$ has mean $m_{0,q}^{(ul,1)}$ and variance $m_{0,q}^{(ul,1)}+\sigma^2_{th}$, and $\boldsymbol{U^{(1)}_1}$ has mean $m_{1,q}^{(ul,1)}$ and variance $m_{1,q}^{(ul,1)}+\sigma^2_{th}$. Also note that $m_{0,q}^{(ul,1)}=E_0+E^{(I)}_q$ and $m_{1,q}^{(ul,1)}=E_1+E^{(I)}_q$, where $E^{(I)}_q=\frac{\eta}{hf}P_cT_c\beta_q^{(I)}$. Therefore, MAI imposes similar effect to additive noise.
  }}
\subsubsection{Conditional CER Analysis for the $i$-th Hop (i=2,...,N+1)} 
 For the remaining hops MAI {does not} affect the chip detection process, but underwater turbulence as well as different noises do affect the system performance and cause errors on the chip detection processes. Similar to \eqref{23} we can express the $q$th chip average received counts of the $i$th hop conditioned on transmitting chips ``OFF" and ``ON" respectively as;
 \begin{subequations}
  \begin{align}
  & m_{0,q}^{(i)}=(\gamma_d+\gamma_b)T_c,\\
   & m_{1,q}^{(i)}=\frac{\eta}{hf}P_cL^{(i)}T_c\tilde{h}^{(i)}+(\gamma_d+\gamma_b)T_c,
  \end{align}
  \end{subequations}
  where $L^{(i)}$ is the $i$th hop channel loss. Consequently, $q$th chip conditional error probabilities of the $i$th hop can be determined { using Gaussian approximation} as{\cite{einarsson2008principles}};
\begin{align} \label{p_{ce-i}}
  & P^{(q)}_{ce-i}(0|1,l,\vec{\alpha},\tilde{h}^{(i)})=P^{(q)}_{ce-i}(1|0,l,\vec{\alpha},\tilde{h}^{(i)})\nonumber\\
  &~~~~~~~~~~~=Q\left(\frac{m_{1,q}^{(i)}-m_{0,q}^{(i)}}{\sqrt{m_{1,q}^{(i)}+\sigma^2_{th}}+\sqrt{m_{0,q}^{(i)}+\sigma^2_{th}}}\right).
  \end{align}
  Note that \eqref{p-ce-1-new} and \eqref{p_{ce-i}} can be applied to \eqref{p_{ce-e2e}} to achieve the end-to-end conditional CERs, while inserting \eqref{p_{ce-e2e}} in \eqref{P_be} results into conditional BERs. Moreover, $P_{be}(\overline{b_0^{(1)}}|b_0^{(1)},l,\vec{\alpha})$ can be determined by averaging \eqref{P_be} over $\bar{H}$ as \eqref{averaging}.
   Here, for uplink transmission the conditional BER equations in \eqref{P_be} can involve maximum of $N+W+1$ fading coefficients as $\bar{H}=(\tilde{h}^{(1)}_{11},\beta_1^{(I)},...,\beta_W^{(I)},\tilde{h}^{(2)},\tilde{h}^{(3)},...,\tilde{h}^{(N+1)})$. For this case joint PDF of $f(\bar{H})$ in \eqref{averaging} simplifies to $f(\bar{H})=f(\tilde{h}^{(1)}_{11})\times\prod_{q=1}^{W} f(\beta_q^{(I)})\times\prod_{i=2}^{N+1}f(\tilde{h}^{(i)})$, due to the independency of fading coefficients. In fact the size of $\bar{H}$ depends on $\vec{\alpha}$ and for each {$\vec{\alpha}$} Eq. \eqref{averaging} can be evaluated through a $J$-dimensional integral, where $J\in\{N+1,N+2,...,N+W+1\}$.\footnote{Note that each $i$th hop ($i=2,3,..,N+1$) introduces a fading coefficient of $\tilde{h}^{(i)}$, while the first hop presents at least one fading coefficient of $\tilde{h}^{(1)}_{11}$. Moreover, the first hop introduces a new fading coefficient of $\beta_q^{(I)}$ for each chip mark $q=1,2,...,W$, as demonstrated in Section III-A1. Depending on $\vec{\alpha}$, $\bar{H}$ has a minimum dimension of $N+1$ (e.g., when $\vec{\alpha}=(0,0,0)$ and all $\beta_q^{(I)}$s are zero) and maximum dimension of $N+W+1$ (e.g., when $\vec{\alpha}=(1,2,3)$ and each chip experiences a new fading coefficient of $\beta_q^{(I)}$).}
  \subsection{Downlink BER Analysis}
  As it has been shown in \cite{ghaffari2008wireless}, MAI of the synchronous downlink  transmission can be eliminated when the number of concurrent users satisfy the following condition;
  \begin{align} \label{bound}
  M<\frac{F}{W^2}+1.
  \end{align}
   In other words, OOC
  codes would not overlap and their internal multiplication, $<C^{(i)}.C^{(j)}>$,
is equal to zero for $i\neq j$. Therefore, $q$th pulsed mark chip conditional error probabilities of the first hop during downlink transmission can be determined similar to those of intermediate hops (in the previous subsection) as;
  \begin{align} \label{p^{dl}_{ce-(1)}}
      & P^{(q,dl)}_{ce-1}(0|1,\tilde{h}^{(1)}_{11})=P^{(q,dl)}_{ce-1}(1|0,\tilde{h}^{(1)}_{11})\nonumber\\
      &~~~~~~~~~~~=Q\left(\!\frac{m_{1,q}^{(dl,1)}-m_{0,q}^{(dl,1)}}{\sqrt{m_{1,q}^{(dl,1)}\!\!+\!\sigma^2_{th}}\!+\!\sqrt{m_{0,q}^{(dl,1)}\!\!+\!\sigma^2_{th}}}\right),
      \end{align}
where ${m_{0,q}^{(dl,1)}}$ and ${m_{1,q}^{(dl,1)}}$ are the $q$th chip average received counts of the first hop conditioned on transmitting chips ``OFF" and ``ON", and are respectively defined as;
\begin{subequations}
  \begin{align}
  & {m_{0,q}^{(dl,1)}}=(\gamma_d+\gamma_b)T_c,\\
   & {m_{1,q}^{(dl,1)}}=\frac{\eta}{hf}P_cL_{1,1}^{(1)}T_c\tilde{h}^{(1)}_{11}+(\gamma_d+\gamma_b)T_c.
  \end{align}
  \end{subequations}
 Conditional CER for each of the other hops is exactly that of uplink transmission in \eqref{p_{ce-i}}. Eventually, relying on \eqref{p_{ce-i}} and \eqref{p^{dl}_{ce-(1)}} the end-to-end conditional CER and BER of downlink transmission can be evaluated similar to \eqref{p_{ce-e2e}} and \eqref{P_be}, respectively. Furthermore, $P_{be}(\overline{b_0^{(1)}}|b_0^{(1)})$ can be determined by averaging \eqref{P_be} over $\bar{H}$ as $P_{be}(\overline{b_0^{(1)}}|b_0^{(1)})=\int_{\bar{H}}P_{be}(\overline{b_0^{(1)}}|b_0^{(1)},\bar{H})f(\bar{H})d\bar{H}$. {It is} obvious that for downlink transmission $\bar{H}$ is defined as a $(N+1)$-dimensional vector as $\bar{H}=(\tilde{h}^{(1)}_{11},\tilde{h}^{(2)},\tilde{h}^{(3)},...,\tilde{h}^{(N+1)})$, and joint PDF of $f(\bar{H})$ simplifies to $f(\bar{H})=f(\tilde{h}^{(1)}_{11})\times\prod_{i=2}^{N+1}f(\tilde{h}^{(i)})$, due to the independency of fading coefficients.
\subsection{Single-User Multi-Hop Point-to-Point UWOC Link}
As a special case, the BER performance of a single-user relay-assisted point-to-point UWOC link with bit-detect-and forward strategy can easily be achieved from those obtained in Sections III-A and III-B, by replacing $M=1$, $F=W=1$, and $T_c=T_b$. In this case based on \eqref{Pr(alpha,l)}-\eqref{p_{ce-e2e}}, the end-to-end bit error probability can be calculated as;
   \begin{align} \label{p_{be-e2e-p2p}}
& P_{be-p2p}(\overline{b_0}|b_0)=\nonumber\\
  &~~~~\int_{\bar{H}}\left(1-\prod_{i=1}^{N+1}\left[1-P_{be-i}(\overline{b_0}|b_0,\tilde{h}^{(i)})\right]\right)f(\bar{H})d\bar{H},
   \end{align}
in which $b_0$ is the $0$th time slot transmitted bit. $\bar{H}$ and $f(\bar{H})$ are defined as $\bar{H}=(\tilde{h}^{(1)},\tilde{h}^{(2)},...,\tilde{h}^{(N+1)})$ and $f(\bar{H})=\prod_{i=1}^{N+1}f(\tilde{h}^{(i)})$, respectively. $P_{be-i}(\overline{b_0}|b_0,\tilde{h}^{(i)})$ is the $i$th hop bit error probability, conditioned on fading coefficient of $\tilde{h}^{(i)}$ and transmitting bit $b_0$ at the $0$th time slot, and can be calculated using Gaussian approximation as{\footnote{It is worth mentioning that all Eqs. \eqref{p_{ce-i}}, \eqref{p^{dl}_{ce-(1)}} and \eqref{P_{be_i}} can also directly be derived from \eqref{p-ce-1-new}, by replacing $l=0$, $\vec{\alpha}=(0,...,0)$, $\beta^{(I)}_q=0$ and $E_q^{(I)}=0$.}};
\begin{align} \label{P_{be_i}}
& P_{be-i}(1|0,\tilde{h}^{(i)})=P_{be-i}(0|1,\tilde{h}^{(i)})\nonumber\\
&~~~~~~~~~~~=Q\!\!\left(\!\frac{m_{1}^{(i)}-m_{0}^{(i)}}{\sqrt{m_{1}^{(i)}\!\!+\!\sigma^2_{th}}\!+\!\sqrt{m_{0}^{(i)}\!\!+\!\sigma^2_{th}}}\right),
\end{align}
where  $m_{0}^{(i)}$ and $m_{1}^{(i)}$ are the mean received counts of the $i$th hop when bits ``$0$" and ``$1$" are transmitted, and are respectively characterized as follows;
\begin{subequations}
  \begin{align}
  & {m_{0}^{(i)}}=(\gamma_d+\gamma_b)T_b,\\
   & {m_{1}^{(i)}}=\frac{\eta}{hf}P^{(i)}_bL^{(i)}T_b\tilde{h}^{(i)}+(\gamma_d+\gamma_b)T_b,
  \end{align}
  \end{subequations}
  in which $P^{(i)}_b$ and $L^{(i)}$ are respectively the transmitted power per bit and channel loss of the $i$th hop. It can be shown that due to the independency of fading coefficients in $\bar{H}$, \eqref{p_{be-e2e-p2p}} simplifies to;
  \begin{align} \label{p_{be-e2e-p2p-averaged}}
    & P_{be-p2p}(\overline{b_0}|b_0)=1-\prod_{i=1}^{N+1}\left[1-P_{be-i}(\overline{b_0}|b_0)\right],
     \end{align}
  where $P_{be-i}(\overline{b_0}|b_0)$ is the $i$th hop BER conditioned on transmitting bit $b_0$ at the $0$th time slot which can be evaluated as $P_{be-i}(\overline{b_0}|b_0)=\int_{\tilde{h}^{(i)}}P_{be-i}(\overline{b_0}|b_0,\tilde{h}^{(i)})f(\tilde{h}^{(i)})d\tilde{h}^{(i)}$. Eq. \eqref{p_{be-e2e-p2p-averaged}} indicates that the average BER in this case can be obtained through the combination of one-dimensional integrals, while it was obtained through $J$-dimensional (or $(N+1)$-dimensional) integral for uplink (or downlink) transmission as we saw in Section III-A (or III-B).

{As we demonstrated the conditional chip/bit error probability of each hop can be considered as a building block for evaluating the end-to-end BERs of each uplink, downlink or single-user transmissions. This error probability can be obtained using either photon-counting method \cite{einarsson2008principles,jamali2015performanceMIMO} or general classical methods in digital communications, similar to \cite{navidpour2007ber,safari2008relay,jamali2015ber}. The second approach approximates the optical signal as a constant coefficient and all of the noise components with an additive Gaussian distributed RV, while photon-counting method, which we used in this paper, is a more accurate method since it considers the incoming optical signal with Poisson distribution. However, it has been shown in \cite{jamali2015performanceMIMO} that both of the above mentioned methods provide relatively the same results. On the other hand, different photon-counting methods exist, e.g., saddle-point and Gaussian approximations \cite{einarsson2008principles}. Saddle-point is a sufficiently accurate method, but involves some complex nonlinear equations and hence its evaluation may be time-consuming. However, in \cite{jamali2015performanceMIMO} the BER of UWOC systems is obtained using both saddle-point and Gaussian approximations and it has been concluded that Gaussian approximation provides relatively similar results to saddle-point approximation. Therefore, in this paper we used Gaussian approximation as a relatively accurate and fast photon-counting method. Moreover, similar to \cite{navidpour2007ber,jamali2015ber} we can use Gauss-Hermite quadrature formula \cite{abramowitz1964handbook} to approximate $J$-dimensional integrals with $J$-dimensional finite series.}
\section{Numerical Results}
In this section, we investigate BER performance of the proposed system for different scenarios. To do so, we consider an underwater network established in \textit{clear ocean}. For this water type parameters $a$, $b$, and $c$ are $0.114$ ${\rm m^{-1}}$, $0.037$ ${\rm m^{-1}}$, and $0.151$ ${\rm m^{-1}}$, respectively. We assume that all users are equidistant to the destination (or to the first relay in the relay-assisted configuration). Further, we assume that when $N$ intermediate relays are employed, all of the hops have the same link length of $d_0/(N+1)$ {m} and therefore the same channel loss and fading log-amplitude variance. To evaluate the channel loss of each hop, we have simulated the channel impulse response based on MC approach similar to \cite{tang2014impulse,cox2012simulation}. 
{Moreover, to obtain fading coefficient of each hop and dependency of these coefficients to the link range we consider specific values for the water turbulence parameters and numerically calculate the scintillation index for various link ranges similar to \cite{korotkova2012light}.}
 Table I displays some important parameters of the system for the channel simulation as well as noise characterization.
 
 {As we mentioned in Section II-B, from the practical aspect of view it is more preferable to use diffusive links for the first and/or last hop(s) to reduce the pointing and tracking difficulties in these hops. In order to include both collimated and diffusive links, we first deal with collimated links developed by narrow laser beams. Then, at the remaining of this section, we investigate the effect of diffusive beams on the system BER and also on the validity of our assumption in Eq. \eqref{compare}.} Table II summarizes the simulation results for the channel loss and scintillation index in several link ranges with the source full beam divergence angle of $\theta_{div}=0.02^0$ and the chip duration time of $T_c=10$ {ns}.
\begin{table}
  \centering
\caption{Some of the important parameters for the channel simulation and noise characterization.}
  \begin{tabular}{>{\centering\arraybackslash}p{1.65in}>{\centering\arraybackslash}p{0.28in}>{\centering\arraybackslash}p{0.8in}}
\toprule[1pt]
   Coefficient & $\!\!$Symbol & Value\\ [0.2ex] 
   \toprule[1pt]
  Half angle field of view & $FOV$ &  ${40}^0$ \\ \hline 
     Receiver aperture diameter & $D_0$ & $20$ {cm} \\ \hline 
     Source wavelength & $\lambda $ & $532$ {nm} \\ \hline 
     Water refractive index & $n$ & $1.331$ \\ \hline 
     Photon weight threshold at the receiver & $w_{th}$ & ${10}^{-6}$ \\ \hline 
     Quantum efficiency & $\eta $ & $0.8$ \\ \hline 
Equivalent temperature& $T_e$& $290$ {K} \\ \hline 
       Load resistance & $R_L$& $100$ $\Omega $\\ \hline
         Dark current & $I_{dc}$ & $1.226\times {10}^{-9}$ {A} \\ \hline
         Background mean count rate &$\gamma_b$& $1.206\times10^{10}$ {1/s} \\ \hline
         Rate of dissipation of mean-square temperature & $\chi_T$  & ${{10}^{-7}}$ ${\rm {{K^2}/{s}}}$ \\ \hline
         Rate of dissipation of turbulent kinetic energy per unit mass of fluid & $\varepsilon $ & $5\times{{10}^{-5}}$ ${\rm {m^2}/{s^3}}$ \\ \hline
         Relative strength of temperature and salinity fluctuations & $w$ & $-3.5$ \\ \hline
         Normalized parameters of the Gaussian beam & $(\Lambda,\Theta)$ & $(0,1)$ \\ \toprule[1pt]
  \end{tabular}
  \vspace{-0.1in}
  \end{table}
  \begin{table}
    \centering
  \caption{Channel loss and scintillation index for several link ranges with $\theta_{div}=0.02^0$ and $T_c=10~\rm{ns}$.}
    \begin{tabular}{||>{\centering\arraybackslash}p{0.4in}|>{\centering\arraybackslash}p{0.7in}|>{\centering\arraybackslash}p{0.7in}|>{\centering\arraybackslash}p{0.8in}||} \hline
      Link range & Channel loss coefficient, $L$ & Scintillation index, $\sigma^2_I$& Log-amplitude variance, $\sigma^2_X$\\ [0.5ex] 
     \hline \hline
 $90$ {m} & $3.99\times10^{-7}$ & $0.9738$ & $0.17$ \\ \hline 
 $70$ {m} & $7.812\times10^{-6}$ & $0.616$ & $0.12$ \\ \hline 
   $45$ {m} & $3.135\times10^{-4}$ & $0.271$ & $0.06$ \\ \hline 
    $30$ {m} & $3.1\times10^{-3} $ & $0.1248$ & $0.029$ \\ \hline 
   $22.5$ {m} & $9.4\times10^{-3} $ & $0.071$ & $0.017$ \\ \hline 
   $18$ {m} & $18.2\times10^{-3} $ & $0.0452$ & $0.011$ \\ \hline 
    \end{tabular}
    \vspace{-0.15in}
    \end{table}
    
In our numerical results we use OOC signature with the following parameters, namely $(F,W,\lambda_a,\lambda_c)=(50,3,1,1)$. Table III summarizes the results of the interference characterization for $M=5$ simultaneous users, each with the aforementioned OOC parameters. First column presents the possible number of interferences on the desired user while the second column indicates probability of having $l$ interferers, defined as $P_l{(l)}=\binom{M-1}{l}\left(\frac{W^2}{2F}\right)^l\left(1-\frac{W^2}{2F}\right)^{M-1-l}$. As it can be observed, larger number of interferers occur with less probability. Third column is the total number of possible interference patterns $\vec{\alpha}$, when $l$ interference occurs on the $W$ chips of the desired user. In other words, this column is the number of $\vec{\alpha}$s satisfying condition $\sum_{q=1}^{W}{{\alpha}_q}=l$, which can be obtained as $\binom{l+W-1}{W-1}$. Fourth column illustrates possible interference patterns for each $l$, while probability of each pattern conditioned on having $l$ interferers, $P_{\vec{\alpha}}(\vec{\alpha}|l)=\frac{l!}{W^l\prod_{q=1}^{W}(\alpha_q)!}$, is determined in the fifth column. Note that different users' interference is more likely to occur on the different chips instead of occurring on a certain chip, e.g., interference pattern of $\vec{\alpha}=(0,0,3)$ is less probable than $\vec{\alpha}=(1,1,1)$. Sixth column represents number of similar $\vec{\alpha}$s for each row of the fourth column, i.e., number of $\vec{\alpha}$s that have the same components as the interference pattern of the fourth column. For example, we have $5$ interference patterns similar to $(0,1,3)$, namely $(0,3,1)$, $(1,0,3)$, $(3,0,1)$, $(1,3,0)$ and $(3,1,0)$. Each of these similar patterns has the same conditional probability of occurrence as the fifth column. Moreover, dimension of $\bar{H}$ in \eqref{averaging} for uplink transmission is summarized in the seventh column. We should emphasize that crosscorrelation constraint of $\lambda_c=1$ imposes that each interfering user can make interference maximum on a one chip of the desired user \cite{salehi1989code}. For example $\vec{\alpha}=(1,1,1)$ indicates three different interferers, each interfering on a certain chip of the desired user; resulting into three new fading coefficients $\beta_1^{(I)}$, $\beta_2^{(I)}$ and $\beta_3^{(I)}$.
\begin{table}
\centering
\caption{Interference characterization for $M=5$ concurrent users with OOC parameters of $(F,W,\lambda_a,\lambda_c)=(50,3,1,1)$.}
\begin{tabular}{||>{\centering\arraybackslash} m{0.05in}|>{\centering\arraybackslash}p{0.3in}|>{\centering\arraybackslash}p{0.35in}||>{\centering\arraybackslash}p{0.35in}|>{\centering\arraybackslash}p{0.3in}|>{\centering\arraybackslash}p{0.35in}|>{\centering\arraybackslash}p{0.5in}||}
 \toprule[1.5pt] \hline
$l$ & $P_l(l)$ & Total \# of $\vec{\alpha}$s, $\binom{l+2}{2}$  & $\vec{\alpha}$  & $P_{\vec{\alpha}}(\vec{\alpha}|l)$ & \# of similar $\vec{\alpha}$s & Dimension of $\bar{H}$ in \eqref{averaging} \\ \hline \hline
                 $0$ &         $0.6857$          &        $1$           & $(0,0,0)$ & $1$ & $0$  & $N+1$ \\ \hline
              $1$    &   $0.2713$                &     $3$              & $(0,0,1)$  & $1/3$ & $2$   & $N+2$ \\ \hline
\multirow{2}{*}{$2$} & \multirow{2}{*}{$0.0402$} & \multirow{2}{*}{$6$} & $(0,0,2)$ & $1/9$ & $2$  & $N+2$ \\ \cline{4-7} 
                  &                   &                   & $(0,1,1)$ & $2/9$ & $2$  & $N+3$ \\ \hline
\multirow{3}{*}{$3$} & \multirow{3}{*}{$0.0027$} & \multirow{3}{*}{$10$} & $(0,0,3)$ & $1/27$ & $2$  & $N+2$ \\ \cline{4-7} 
                  &                   &                   & $(0,1,2)$ & $1/9$ & $5$ & $N+3$ \\ \cline{4-7} 
                  &                   &                   & $(1,1,1)$ & $2/9$ & $0$ & $N+4$ \\ \hline
\multirow{5}{*}{$4$} & \multirow{5}{*}{$\begin{matrix}
6.561 \\ \times10^{-5}
\end{matrix}$} & \multirow{5}{*}{$15$} & $(0,0,4)$ & $1/81$ & $2$  & $N+2$ \\ \cline{4-7} 
                  &                   &                   & $(0,1,3)$ & $4/81$ & $5$  & $N+3$ \\ \cline{4-7} 
                  &                   &                   & $(0,2,2)$ & $2/27$ & $2$ & $N+3$ \\ \cline{4-7} 
                  &                   &                   & $(1,1,2)$ & $4/27$ & $2$ & $N+4$ \\ \hline
\end{tabular}
\vspace{-0.15in}
\end{table}

 \begin{figure}
                \centering
                \includegraphics[width=3.6in]{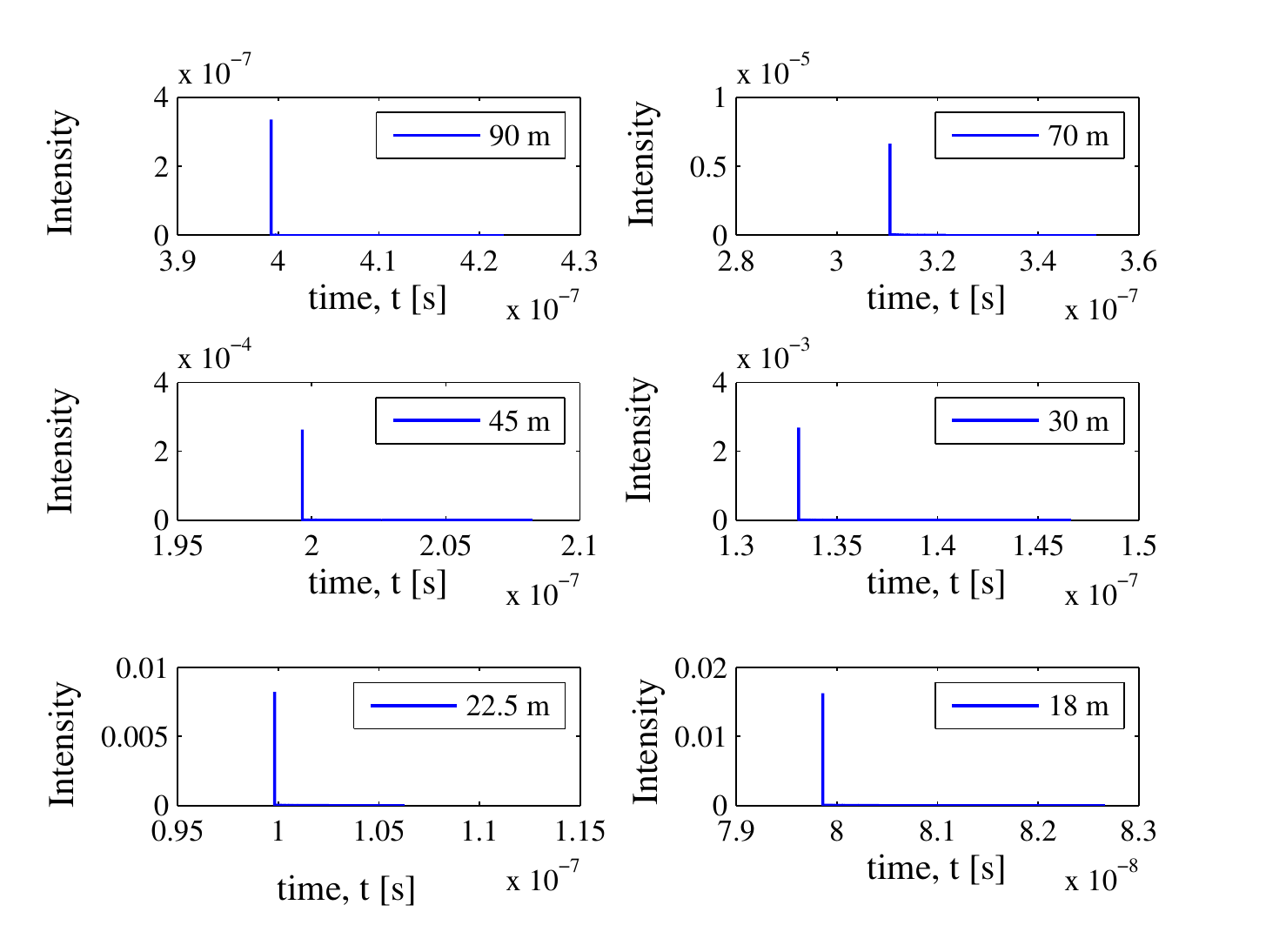}
            \caption{Fading-free impulse response of the clear ocean channel for various link ranges, $d_0=90$ {m}, $70$ {m}, $45$ {m}, $30$ {m}, $22.5$ {m}, and $18$ {m}.}
                        \vspace{-0.15in}
                \end{figure}
Fig. 3 illustrates the fading-free impulse response of the channel for various link ranges, simulated based on MC method. As it can be seen, even for link range of $d_0=90$ {m} the clear ocean channel impulse response can approximately be modeled by an ideal delta function. This behaviour strongly guarantees \eqref{compare}, particularly for $T_c=10$ {ns}.
  Therefore, we consider the chip duration time as $T_c=10$ {ns}, and consequently each user can communicate with data rate of $R_b=2$ {Mbps}.
 
Fig. 4 indicates the scintillation index $\sigma^2_I$ and the corresponding log-amplitude variance $\sigma^2_X$ of the channel in terms of link range, with respect to the turbulence parameters in Table I. Furthermore, effects of increase on these parameters is investigated in this figure. As it can be seen, increases on $\chi_T$ and $w$ results into stronger fading while smaller value for the scintillation index is expected for larger $\varepsilon$.
    \begin{figure}
          \centering
          \includegraphics[width=3.6in]{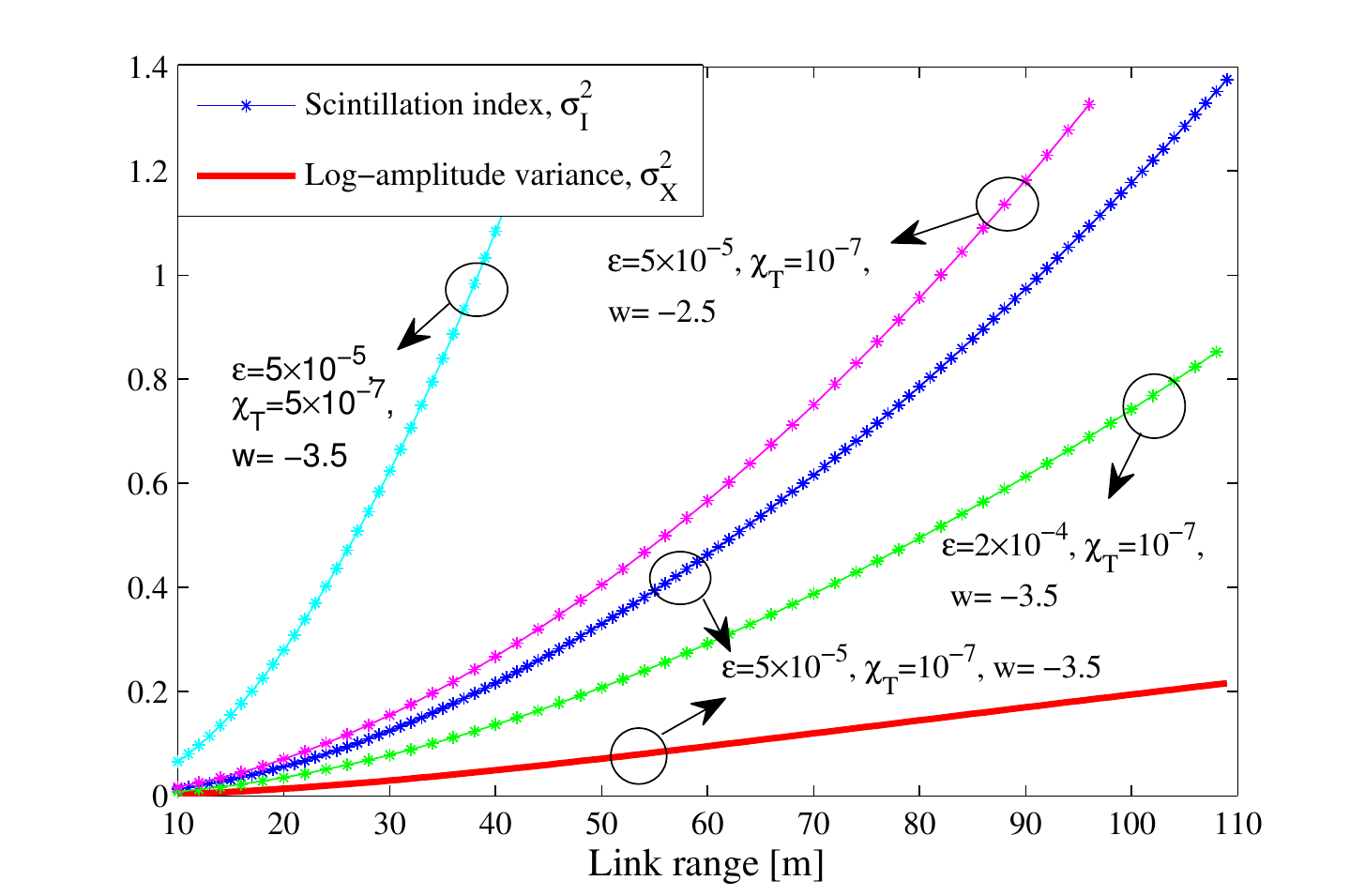}
       \caption{Effect of increasing turbulence parameters $\varepsilon $, $\chi_T$ and $w$ on the scintillation index $\sigma^2_I$.}
                   \vspace{-0.15in}
          \end{figure}
          
\begin{figure}
            \centering
            \includegraphics[width=3.6in]{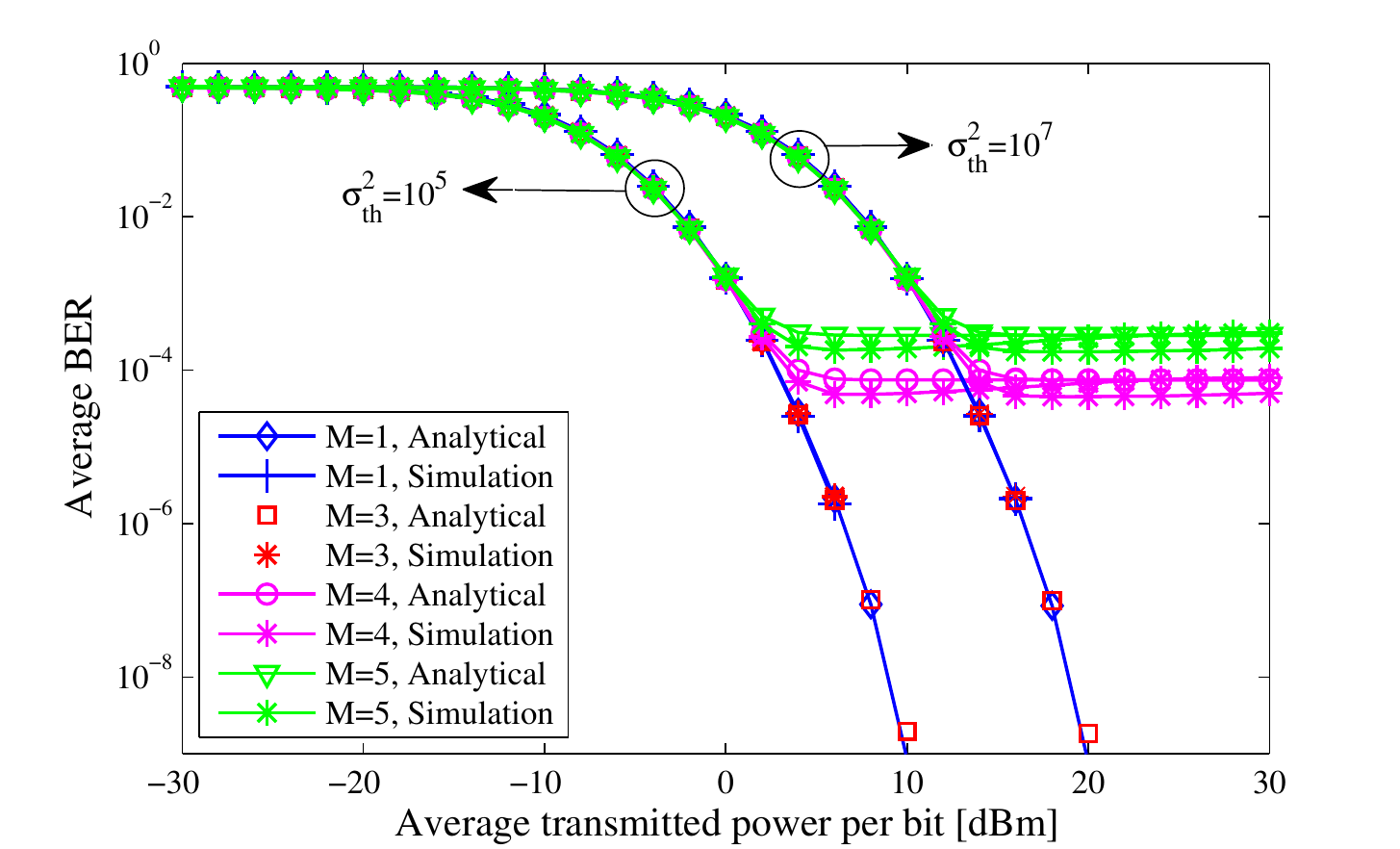}
         \caption{Effect of increasing the number of simultaneous users $M$ and the thermal noise variance $\sigma^2_{th}$ on the uplink performance of a $d_0=70$ {m} clear ocean link without any intermediate relay node.}
 \vspace{-0.15in}
\end{figure}
{In order to perform the numerical simulations we generate $10^{7}$ Bernoulli RVs with PDF of $\Pr(b_k^{(n)})=\frac{1}{2}\delta(b_k^{(n)})+\frac{1}{2}\delta(b_k^{(n)}-1)$ for each user, where $\delta(.)$ is Dirac delta function. Then, we transmit $10^{7}$ Poisson distributed RVs with mean $\frac{\eta}{hf}P_cT_cb_k^{(n)}$ for each chip and multiply these RVs with the link loss and with $10^{7}$ log-normal RVs corresponding to the link fading coefficient. At the receiver side we add the resulted sequence with $10^{7}$ Gaussian distributed RVs that correspond to the receiver thermal noise and with $10^{7}$ Poisson distributed RVs corresponding to the receiver dark current and background noise. Afterward, we compare each of the $10^{7}$ resulted decision variables with an appropriate threshold to detect the received chips. The detected chip of each node would be transmitted via a Poisson distributed RV with mean $\frac{\eta}{hf}P_cT_cC_q^{(i)}$, where $C_q^{(i)}$ is the $q$th detected chip of the $i$th node. Pursuing similar procedures, we can detect the received chips at the destination. Then, if all the chips of a time slot are ``ON" we recognize bit ``$1$" as the transmitted data bit and vice versa. Finally, the BER can be evaluated comparing the transmitted and the decoded data sequences.}

To see effects of increasing the number of simultaneous users on the uplink performance, in Fig. 5 the uplink BER of a $d_0=70$ {m} clear ocean link without any intermediate relay is depicted for $M=1$, $3$, $4$ and $5$, $R_b=2$ {Mbps}, and two different values of the thermal noise variance, namely $\sigma^2_{th}=10^5$ and $10^7$. {As it can be seen, when the number of simultaneous users is smaller or equal to the code weight ($M\leq W$) MAI does not affect the system performance. However, as the number of users violates the code weight, the uplink BER saturates to a constant value. For example, when $M=4$ the main contribution of BER is due to the interference pattern of $\vec{\alpha}=(1,1,1)$. In fact for this interference pattern $P_{be}(1|0)\rightarrow1, P_{be}(0|1)\rightarrow0$ and $P_E\rightarrow\frac{1}{2}P_l(l=3,\vec{\alpha}=(1,1,1))=\frac{1}{2}\times1.62\times10^{-4}=8.1\times10^{-5}$. Note that as the number of users increases the saturation value increases too. Also larger BERs are expected for larger values of the thermal noise variance.}

{However, the decision rule of Eq. \eqref{P_be} appropriately mitigates the bit error probabilities asymmetry for weak to moderate SNR regimes, at high SNR regimes the severe increase on the photoelectrons count of the desired user's chips, caused by high interference power of undesired users, considerably increases the difference between error probabilities for chips ``ON" and ``OFF". In other words, at high SNR regimes where the system is MAI-limited $P^{(q)}_{ce-e2e}(1|0,l,\vec{\alpha},\bar{H})\gg P^{(q)}_{ce-e2e}(0|1,l,\vec{\alpha},\bar{H})$ and hence $P_{be}(1|0,l,\vec{\alpha},\bar{H})\gg P_{be}(0|1,l,\vec{\alpha},\bar{H})$. The aforementioned asymmetry is mainly due to the simple threshold value of the receiver that only takes into account the transmitted power of the desired user and its channel condition. In general, maximum likelihood (ML) hypothesis should be considered to derive the optimum threshold value and hence the optimum receiver structure. Undoubtedly, the exact BER of the OCDMA system with optimum threshold value (which relates to the number of interfering users and their transmitted power and channel condition) would be less than the above mentioned simple receiver's BER. Therefore, the receiver structure adopted in this paper provides a simple decision rule and hence an upper bound on the BER \cite{shalaby1998chip}.}

\begin{figure}
      \centering
      \includegraphics[width=3.6in]{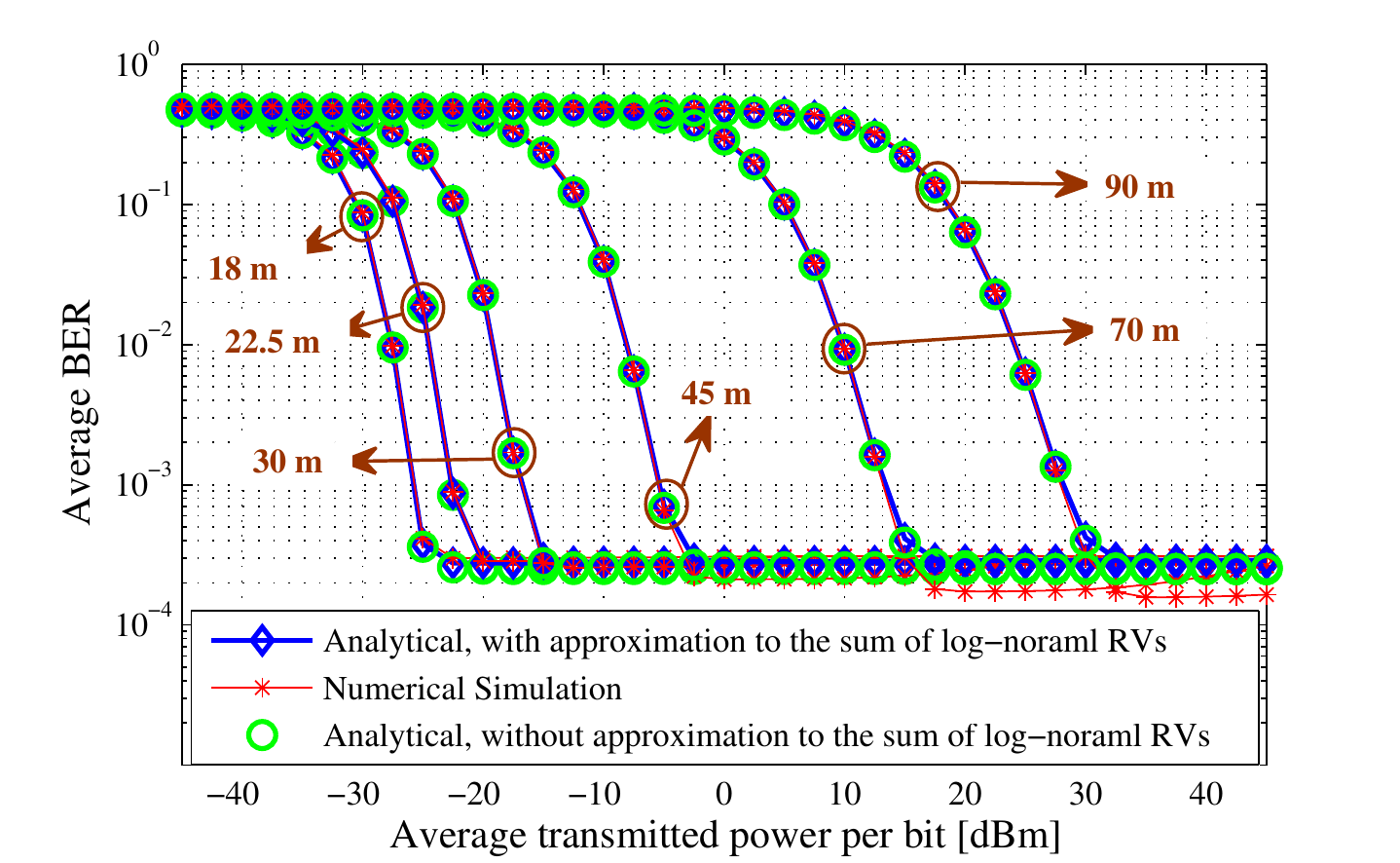}
 \caption{Uplink BER of an OCDMA-based underwater network without any intermediate node. $M=5$, $\sigma^2_{th}=3.12\times10^7$, $R_b=2$ {Mbps}, and various communication ranges $d_0=90$ {m}, $70$ {m}, $45$ {m}, $30$ {m}, $22.5$ {m} and $18$ {m}.}
             \vspace{-0.15in}
      \end{figure}            
Fig. 6 illustrates uplink BER of the underwater network for various link ranges and $M=5$ simultaneous users that transmit OCDMA-based data to a destination directly without any intermediate relay. {Also based on Eq. \eqref{thermal} and values of the parameters in Table I, the thermal noise variance is $\sigma^2_{th}=3.12\times10^7$}. As it can be observed, increasing the link range severely degrades the performance. This is reasonable since the degrading effects of absorption, scattering and fading increase rapidly with distance. Hence, the idea behind multi-hop transmission becomes apparent; divide the long communication distance to shorter ones (by means of intermediate relays) each with much reduced degrading effects of absorption, scattering and fading. {Moreover, the well matches between the analytical and numerical simulation results show the accuracy of our derived expressions. However, a few discrepancy exists at low error rates, mainly due to numerical calculation of multi-dimensional integrals of analytical expressions.
 We also calculated the uplink BER without approximation to the sum of log-normal RVs, using $M$-dimensional integrals. As it can be seen, the well match between the results confirms the accuracy of the mentioned approximation in Eqs. \eqref{sum_log}-\eqref{sigma_z}. In general, this approximation is more beneficial when $M\geq W+1$, since it converts $(M+N)$-dimensional integrals to $(W+1+N)$-dimensional counterparts.}

\begin{figure}
            \centering
            \includegraphics[width=3.6in]{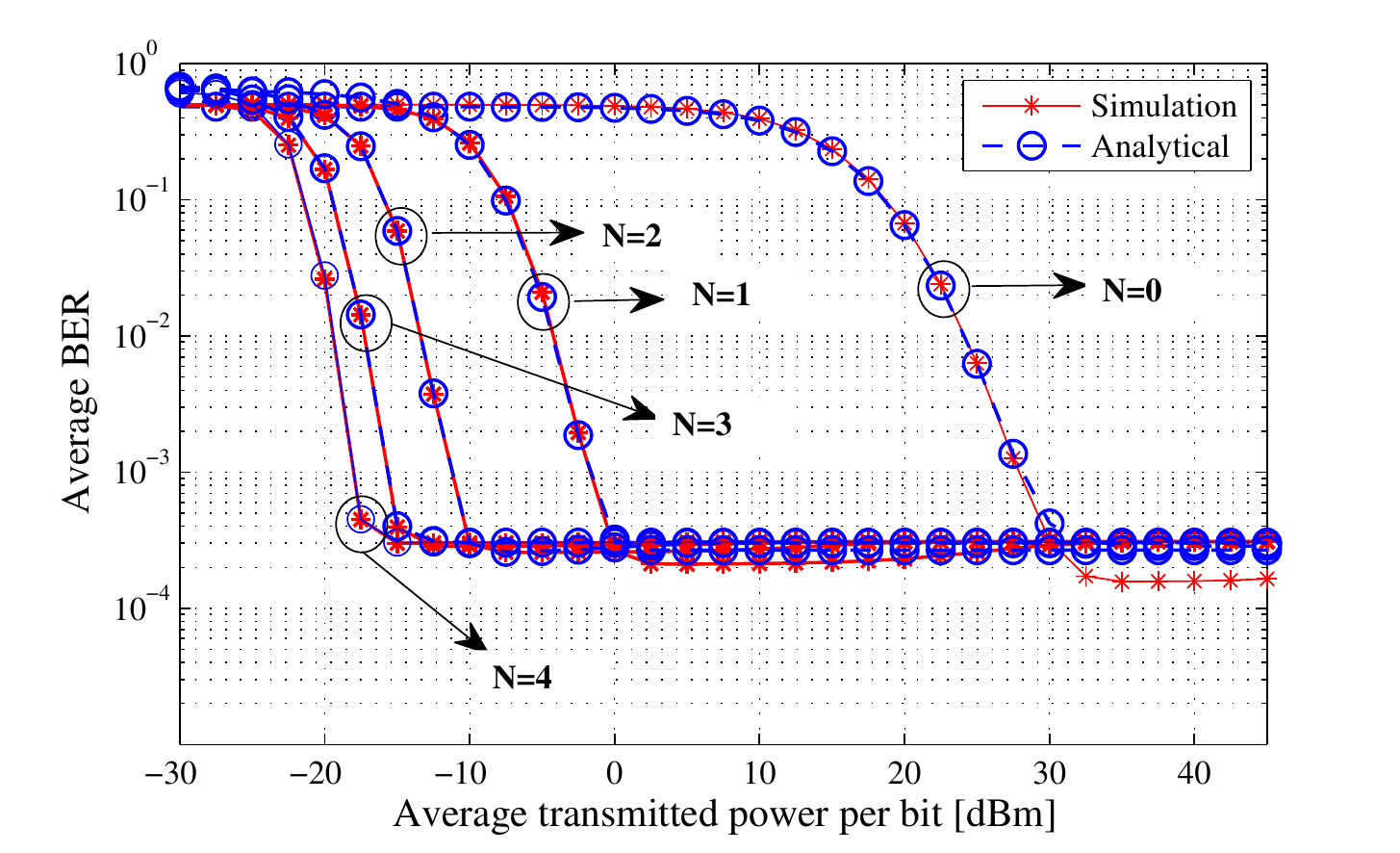}
\caption{Uplink BER of the relay-assisted OCDMA-based underwater network with $M=5$ users and end-to-end distance of $d_0=90$ {m}. $R_b=2$ {Mbps}, $\sigma^2_{th}=3.12\times10^7$ and $N=0$, $1$, $2$, $3$ and $4$.}
            \vspace{-0.15in}
            \end{figure}   
Fig. 7 depicts uplink BER of the relay-assisted UWOC network where $M=5$ OCDMA-based simultaneous users are communicating with a destination located $90$ {m} away in clear ocean. In this scenario, we assume that all of the transmitters transmit their corresponding chips with the same multi-hop chip power of $P_c^{mh}=\frac{2F}{(N+1)W}P_{b,avg}$, where $P_{b,avg}$ is the average transmitted power per bit of each user. As it is shown in Fig. 7, significant performance improvement can be achieved by utilizing intermediate relays, e.g., {$30$} {dB} improvement in the BER of {$10^{-3}$} by employing dual-hop transmission. Comparing Figs. 6 and 7 shows that in multi-hop transmission impairment of each hop is a bottleneck on the overall performance. In other words, a multi-hop link with $N$ intermediate relays and $d_0$ {m} end-to-end distance can approximately achieve the performance of a single-hop link with $d_0/(N+1)$ {m} communication distance and by approximately $10~{\rm log}_{10}(N+1)$ {dB} larger average transmitted power per bit (here $d_0=90$ {m} and $N=0,1,...,4$).
                        
\begin{figure}
            \centering
            \includegraphics[width=3.6in]{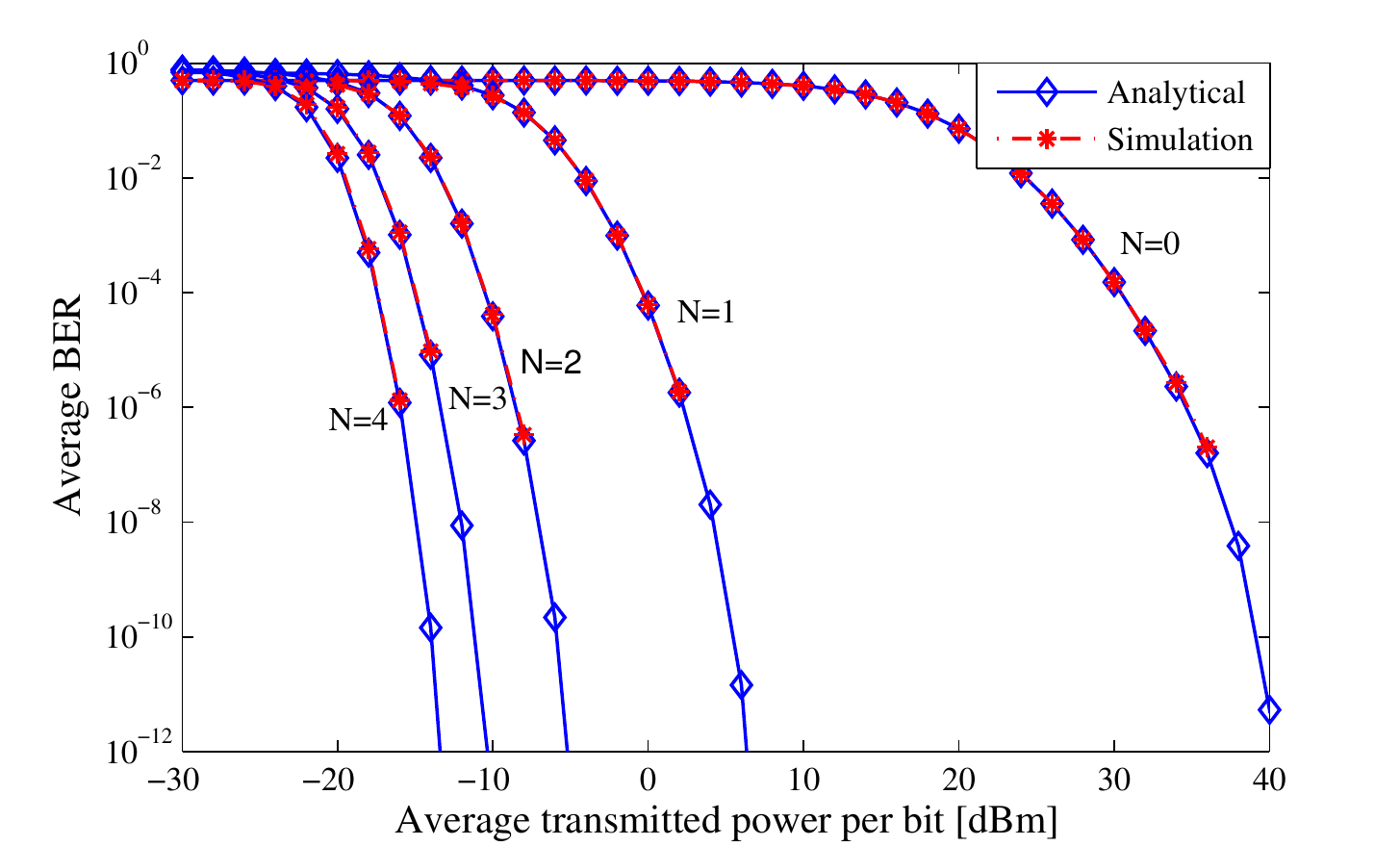}
\caption{Downlink BER of the relay-assisted OCDMA-based underwater network with $M=5$ simultaneous users. $d_0=90$ {m}, $R_b=2$ {Mbps}, $\sigma^2_{th}=3.12\times10^7$, and $N=0$, $1$, $2$, $3$ and $4$.}
            \vspace{-0.15in}
\end{figure} 
Fig. 8 indicates downlink BER of the relay-assisted OCDMA-based UWOC network. For this case OBTS transmits synchronous data to a collection of $M=5$ users (located $90$ {m} away in clear ocean), which satisfies the inequality in \eqref{bound}. Therefore, downlink transmission {does not} suffer from MAI. Consequently, downlink BER can surpass that of uplink {(when $M\geq W+1$)}, as it is apparent by comparing Figs. 7 and 8. Moreover, similar to uplink transmission multi-hop downlink transmission can yield significant performance improvement, e.g., $32.5$ {dB} in the BER of $10^{-6}$ by means of only one intermediate relay.

\begin{figure}
                        \centering
                        \includegraphics[width=3.6in]{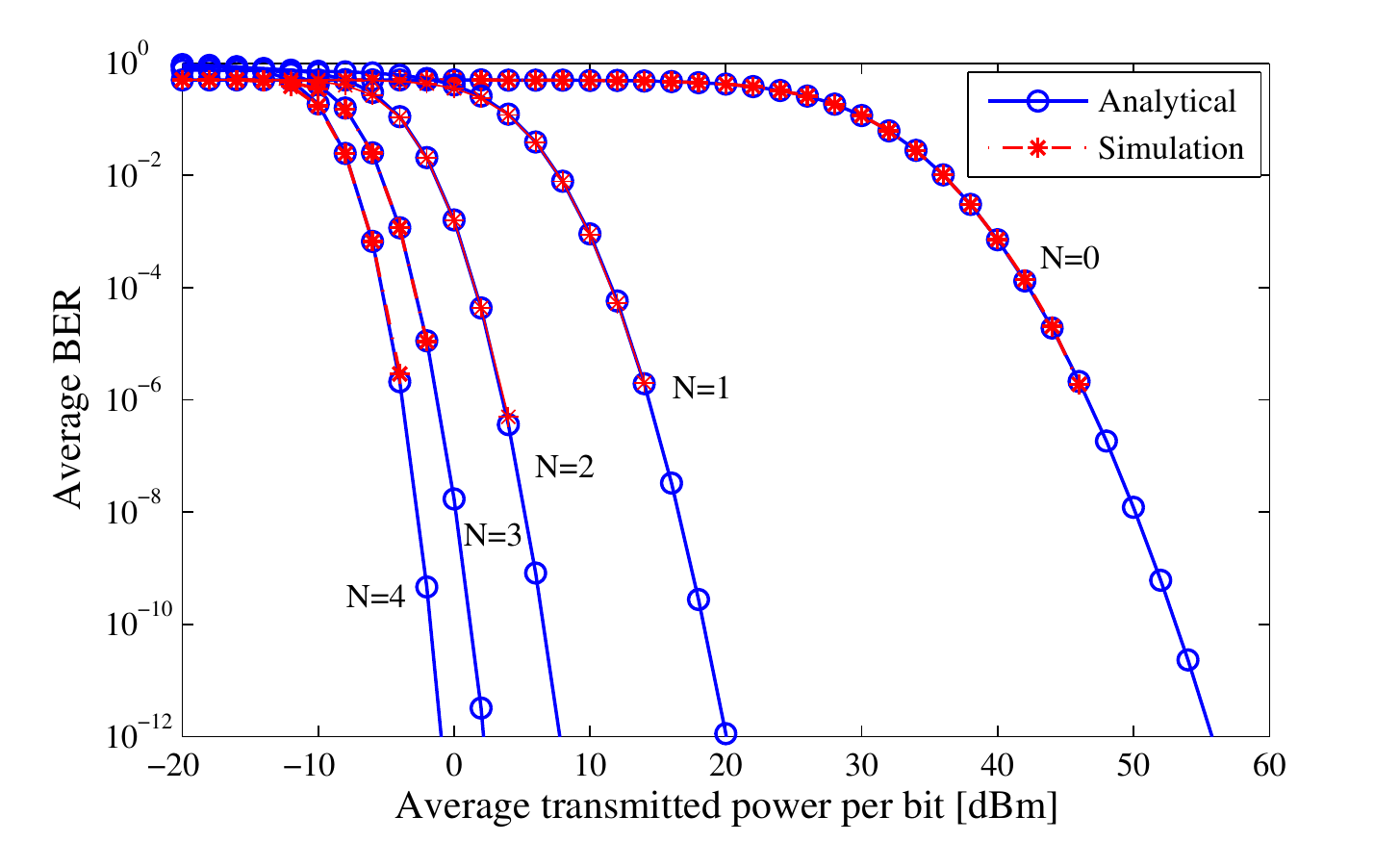}
            \caption{BER performance of a single-user relay-assisted point-to-point UWOC system. $d_0=90$ {m}, $R_b=100$ {Mbps}, $\sigma^2_{th}=3.12\times10^7$, $N=0$, $1$, $2$, $3$ and $4$.}
            \vspace{-0.15in}
\end{figure}
Fig. 9 shows BER performance of a single-user relay-assisted point-to point UWOC system with data rate of $100$ {Mbps} in a $90$ {m} clear ocean link. As it can be seen multi-hop transmission significantly improves the system performance, e.g., $32$ {dB} improvement in the BER of $10^{-6}$ is achieved using a dual-hop transmission. {We should emphasize that the negligible discrepancy between the simulation and analytical results of relay-assisted topology at high BER regimes is caused by the assumption in \eqref{p_{ce-e2e}}, which provides a relatively tight upper bound on the system BER.}
                        
\begin{figure}
            \centering
            \includegraphics[width=3.6in]{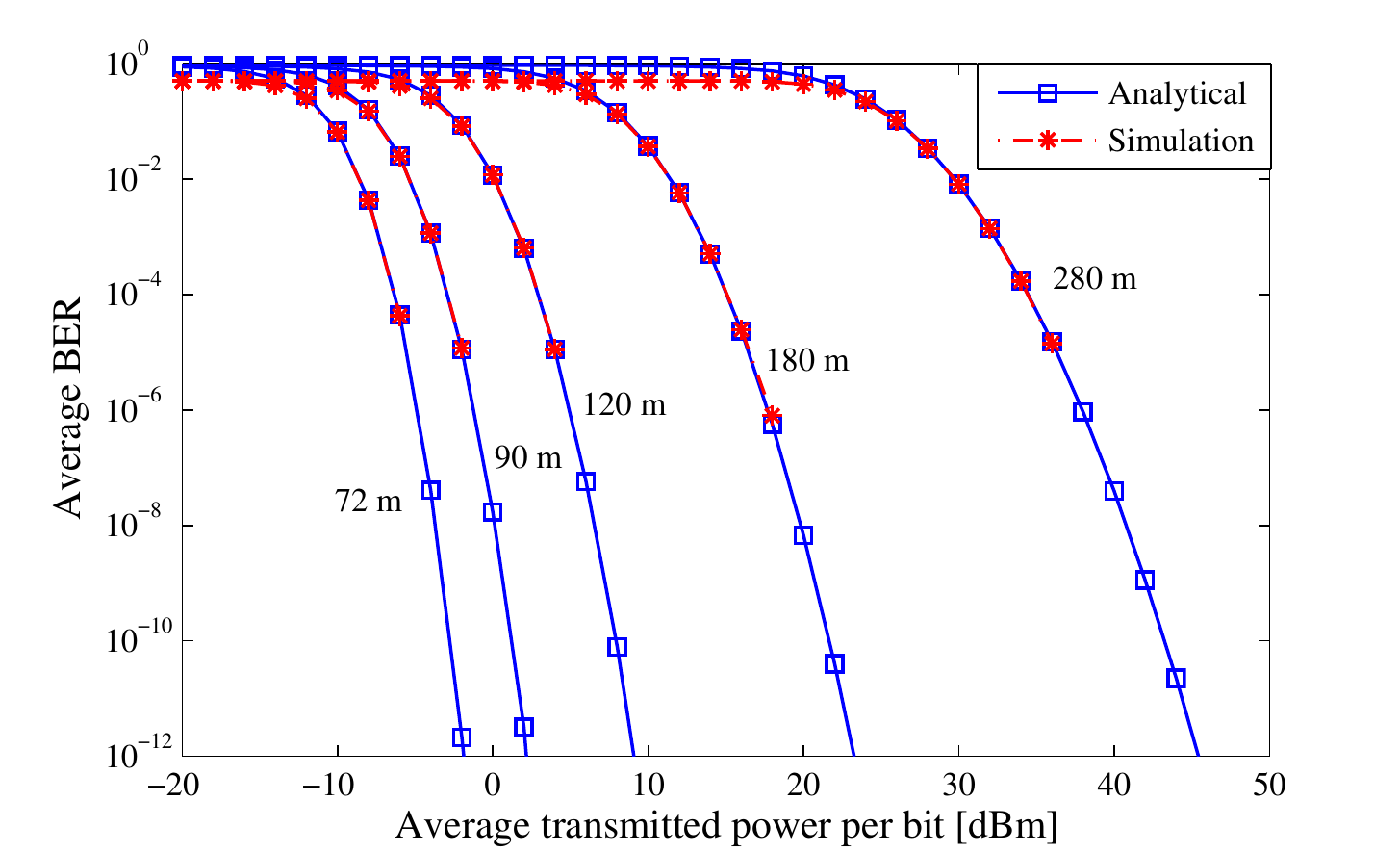}
\caption{Beneficial application of the multi-hop transmission (serial relaying) in expanding the viable communication range; BER of a single-user relay-assisted point-to-point UWOC system with $N=3$ intermediate relays, in several link ranges, i.e., $d_0=280$ {m}, $180$ {m}, $120$ {m}, $90$ {m} and $72$ {m}.}
            \vspace{-0.15in}
\end{figure}
In order to more clearly illustrate the beneficial application of multi-hop transmission in extending the viable communication distance, we consider a single-user relay-assisted point-to-point UWOC system with $N=3$ intermediate relays. Fig. 10 depicts the BER of such a system in several link ranges. As it can bee seen, by means of multi-hop transmission we can attain comparatively large communication distances with realistic transmit powers and acceptable BERs. For instance, we can communicate through a $280$ {m} clear ocean link with an appropriate performance and a realistic average transmitted power per bit. These achievable communication distances substantially differ from the reported distances in the existing literature (i.e., \textit{``typically shorter than $100$ {m}"}). Our numerical results confirm that employment of intermediate relay nodes is of utmost importance for underwater optical communication. {Appropriate number of relays can be obtained using the network designing policies. For example if the system is power constrained, we can first obtain the BER of the system under a set of conditions versus the transmitted power and then we can choose the number of relays such that an acceptable performance achieves using a predetermined total transmitted power. Moreover, the largest link span depends on the system parameters and for a given set of conditions and system parameters it can easily be determined using our derived expressions for the system BER.}

{However, our previous numerical results dealt with collimated laser beams for all hops, at the remaining of this section we investigate the effect of diffusive links.}
 Fig. 11 illustrates the MC simulation results for the fading-free impulse response of UWOC channel with different link ranges, namely $d_0=90$ {m}, $45$ {m}, $30$ {m} and $20$ {m}. Also the transmitter full beam divergence angle is assumed to be $\theta_{div}=5^0$. As it can be seen, even for $90$ {m} clear ocean link the channel spreading time is negligible with respect to the chip duration time. Based on \eqref{loss}, the channel loss coefficients for to the above mentioned link lengths are $5.9224\times10^{-9}$, $1.8575\times10^{-5}$, $2.5194\times10^{-4}$ and $1.6167\times10^{-3}$, respectively.
\begin{figure}
 \centering
\includegraphics[width=3.6in]{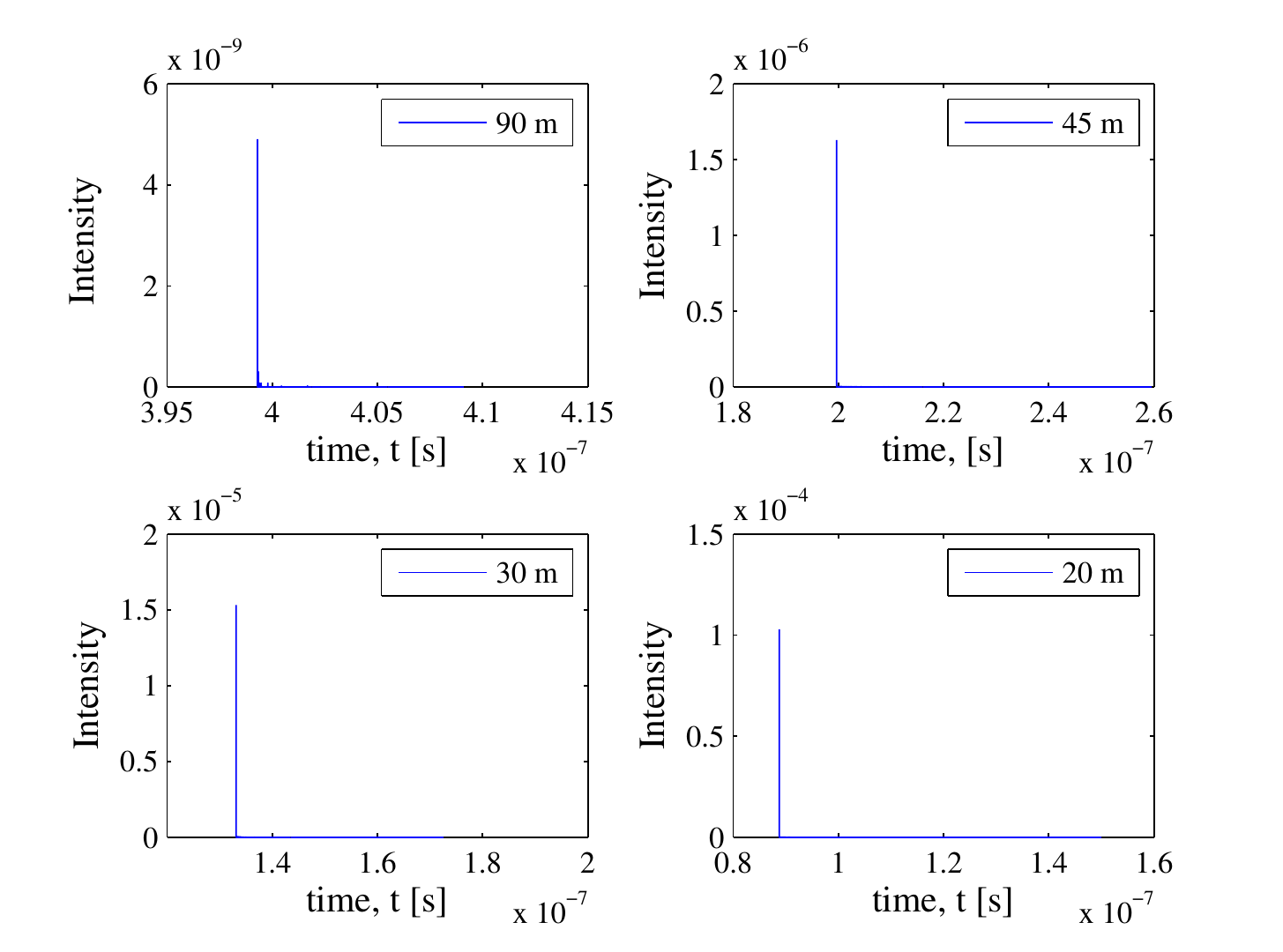}
\caption{Fading-free impulse response of the clear ocean link with transmitter beam divergence angle of $\theta_{div}=5^0$ and various link lengths, namely $d_0=90$ {m}, $45$ {m}, $30$ {m} and $20$ {m}.}
\vspace{-0.15in}
\end{figure}

\begin{figure}
            \centering
            \includegraphics[width=3.6in]{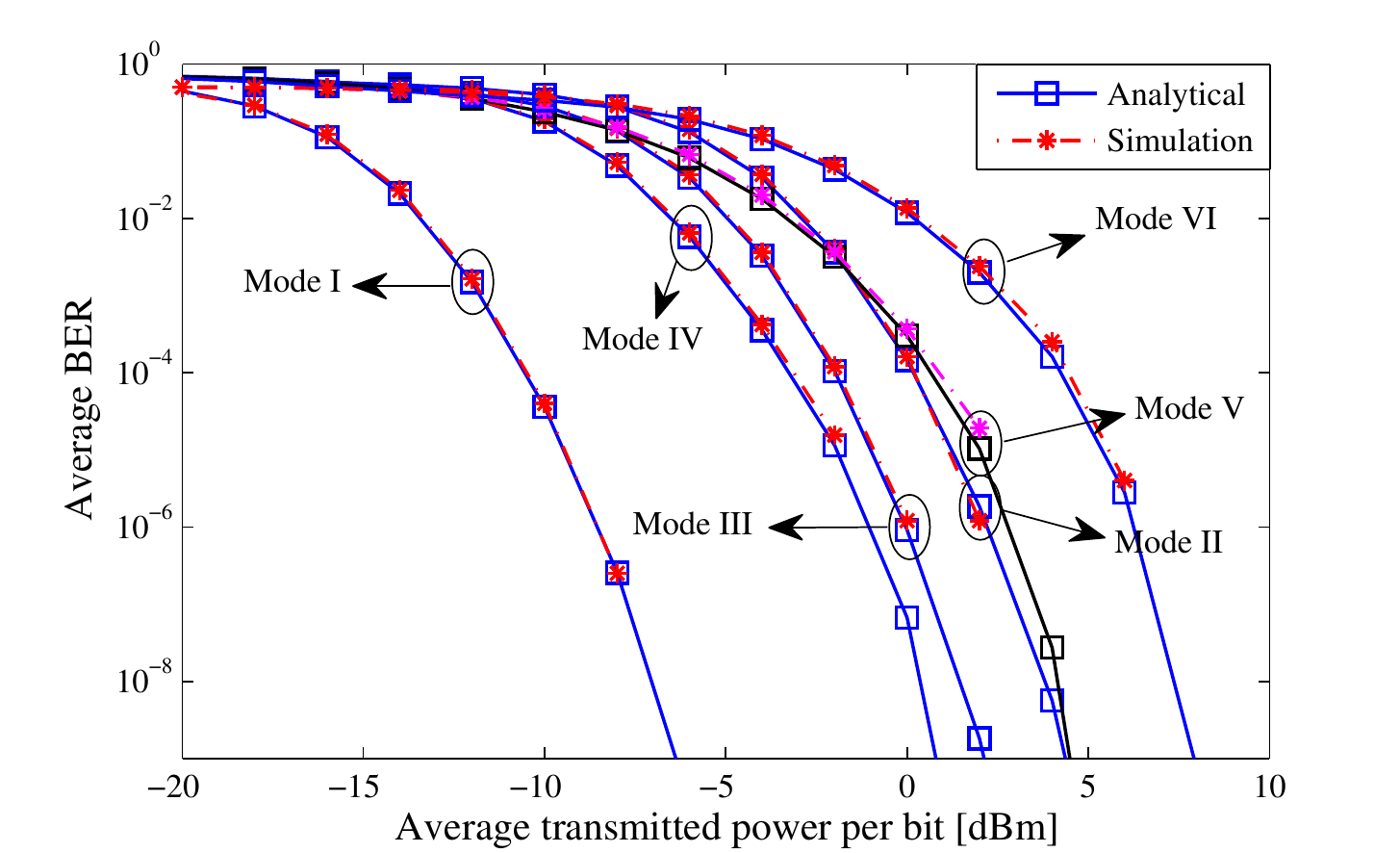}
\caption{Uplink BER of a relay-assisted OCDMA-UWOC network with $M=3$, $N=2$, $d_0=90$ {m} and $\sigma^2_{th}=3.12\times10^7$. The effects of diffusive links for the first and last hops as well as changing the position of intermediate relays are investigated.}
\vspace{-0.15in}
\end{figure}
{
Fig. 12 shows the uplink BER of a relay-assisted OCDMA-UWOC network with $M=3$ users, $N=2$ intermediate relays and $d_0=90$ {m} end-to-end communication distance. In this figure we consider diffusive transmitters with $\theta_{div}=5^0$ for the first and last hops. The transmitter of the second hop (between two relays) is yet assumed to be collimated with $\theta_{div}=0.02^0$. We also investigate the effect of changing the position of intermediate relays on the system BER. for this purpose we define $d_{0,ij}$ as the length of the hop between the nodes $i$th and $j$th. Here, $d_{0,12}+d_{0,23}+d_{0,34}=90$ {m}. We define different modes of operation for the assumed network. Mode I specifies an OCDMA-UWOC network with equal hop lengths as $d_{0,12}=d_{0,23}=d_{0,34}=30$ {m}, same $FOV=40^0$ for all of the receivers and equal divergence angle of $\theta_{div}=0.02^0$ for all of the transmitters. Mode II relates to the same parameters as Mode I and only the transmitters of the first and last hops are assumed diffusive with $\theta_{div}=5^0$. For Modes III, IV, V and VI we assume exactly the same parameters as Mode II, except that the effect of changing the position of intermediate nodes is investigated. Specifically, $d_{0,12}=d_{0,34}={27.5}$ {m} and $d_{0,23}=35$ {m} for Mode III, $d_{0,12}=d_{0,34}=25$ {m} and $d_{0,23}=40$ {m} for Mode IV, $d_{0,12}=d_{0,34}=22.5$ {m} and $d_{0,23}=45$ {m} for Mode V, and $d_{0,12}=d_{0,34}=20$ {m} and $d_{0,23}=50$ {m} for Mode VI. The channel loss coefficient as well as fading statistics for different link rages and various transmitter beam divergence angles are calculated and summarized in Table IV. As it can be seen, increasing the beam divergence angle of the first and last hops' transmitters degrades the system performance; however it is important from the practical point of view. Moreover, our numerical results show that we can achieve a better performance by changing the position of intermediate relays, i.e., there is an optimal place for intermediate relays from the BER viewpoint. For example, Mode IV has approximately $7.5$ {dB} better performance than Mode VI at the BER of $10^{-6}$. It is worth noting that as the total link range and the beam divergence angles of diffusive transmitters increase, the differences between performances of various modes increase too and hence the optimal relay placement achieves more importance.
\begin{table}
    \centering
  \caption{Channel loss and scintillation index for several link ranges corresponding to Fig. 12.}
    \begin{tabular}{||>{\centering\arraybackslash}p{0.35in}|>{\centering\arraybackslash}p{0.5in}|>{\centering\arraybackslash}p{0.68in}|>{\centering\arraybackslash}p{0.48in}|>{\centering\arraybackslash}p{0.4in}||} \hline
      Link range & Full beam divergence angle, $\theta_{div}$ & Channel loss coefficient, $L$ & Scintillation index, $\sigma^2_I$& Log-amplitude variance, $\sigma^2_X$\\ [0.5ex] 
     \hline \hline
 $30$ {m} &$5^0$& $2.5194\times10^{-4}$ &$0.1248$& $0.0294$\\ \hline 
 $35$ {m} &$0.02^0$& $1.4722\times10^{-3}$&$0.1681$ & $0.03884$ \\ \hline 
   $27.5$ {m} &$5^0$& $3.8713\times10^{-4}$ &$0.1054$& $0.02505$  \\ \hline 
    $40$ {m} &$0.02^0$& $6.9336\times10^{-4} $ &$0.2169$& $0.04907$ \\ \hline 
   $25$ {m} &$5^0$& $6.1802\times10^{-4} $ &$0.0873$& $0.02093$\\ \hline 
   $22.5$ {m} &$5^0$& $1.0015\times10^{-3} $ &$0.0712$& $0.0172$ \\ \hline 
   $50$ {m} &$0.02^0$& $1.5282\times10^{-4} $ &$0.3303$& $0.07136$\\ \hline 
   $20$ {m} &$5^0$& $1.6167\times10^{-3} $ &$0.0559$& $0.01361$ \\ \hline 
    \end{tabular}
    \vspace{-0.15in}
    \end{table}
}
\section{Conclusions}
{In this paper, we studied the BER performance of optical code division multiple access (OCDMA)-based underwater wireless networks for both up-and downlink transmissions. We considered all the disturbing effects of the medium to better model the underwater channel. 
To benefit from the incremental dependency of absorption, scattering and fading to the underwater link range, we considered relay-assisted transmission to divide a relatively long communication distance to shorter ones, each with much reduced absorption, scattering and fading effects. For the sake of simplicity, we assumed chip detect-and-forward strategy at the relay nodes. In order to evaluate the chip error probability of each hop and then the end-to-end BERs, we applied Gaussian approximation which is based on photon-counting methods. We found that when the number of simultaneous users is smaller or equal to the code weight ($M\leq W$), multiple access interference (MAI) does not affect the system performance and therefore both up-and downlink transmissions have the same end-to-end BERs. However, as the number of users violates the code weight uplink BER saturates to a constant value; whereas downlink BER does not change, since MAI is removed by synchronous downlink transmission.
 Furthermore, BER of a single-user relay-assisted point-to-point UWOC system obtained as a special case of the proposed relay-assisted OCDMA network. Based on our numerical results multi-hop transmission considerably improves the performance, especially for longer link ranges, e.g., $32.5$ {dB} performance enhancement obtained via only one intermediate relay node for downlink transmission in a $d_0=90$ {m} clear ocean link and at the BER of $10^{-6}$. Also we numerically showed that there could be an optimal place for relay nodes from the BER point of view. Moreover, our simulation results confirmed that the employment of intermediate relay nodes is of utmost importance for underwater wireless optical communication, especially for larger distances.}
 


\begin{IEEEbiography}[{\includegraphics[width=1in,height=1.25in,clip,keepaspectratio]{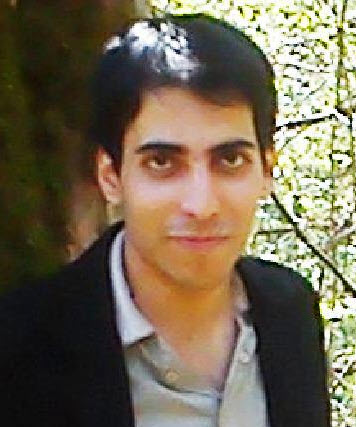}}]{Mohammad Vahid Jamali}
was born in
Talesh, Iran, on May 22, 1991. He received
the B.Sc. degree from K.N. Toosi University of Technology,
Tehran, Iran, in 2013, and the M.Sc. degree from
Sharif University of Technology (SUT), Tehran, Iran, in 2015, both with honors and in Electrical
Engineering.
Since 2013, he has been a member of
the technical staff of the Optical Networks Research
Lab (ONRL) at SUT. 
His general
research interests are in communications theory and
optics with special emphasis on wireless applications. Specific research areas include underwater wireless optical communications, mode-division multiplexing in optical fibers, free-space optics, visible light communications, simultaneous information and power transfer, metamaterials and metasurfaces, and DNA sequencing based on optical computing.
\end{IEEEbiography}
\begin{IEEEbiography}[{\includegraphics[width=1in,height=1.25in,clip,keepaspectratio]{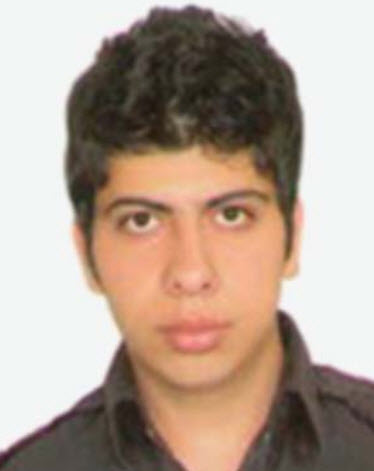}}]{Farhad Akhoundi} received the B.Sc. degree (first-class honor) from Shahid Rajaee Teacher Training University (SRTTU), Tehran, in 2010, and the M.Sc. degree from Sharif University of Technology (SUT), Tehran, Iran, in 2012, both in electrical engineering. From 2012 to 2014, he was a member of the technical staff of the Optical Networks Research Laboratory (ONRL) at SUT. From 2014 to 2015, he was employed by DNA Microarray Analysis Lab (DMA Lab), a knowledge-based company in SUT, working on camera-based DNA microarray scanner. Farhad is currently working toward the Ph.D. degree in the University of Arizona (UA), Tucson, USA. He is also a Full-Time Graduate Research Assistant at College of Optical Sciences, UA. His research interests are in the areas of fiber optics, visible light, and underwater wireless optical multiple access networks. He is also interested in multi-photon microscopy and fluorescence imaging in the biosciences.
\end{IEEEbiography}
\begin{IEEEbiography}[{\includegraphics[width=1in,height=1.25in,clip,keepaspectratio]{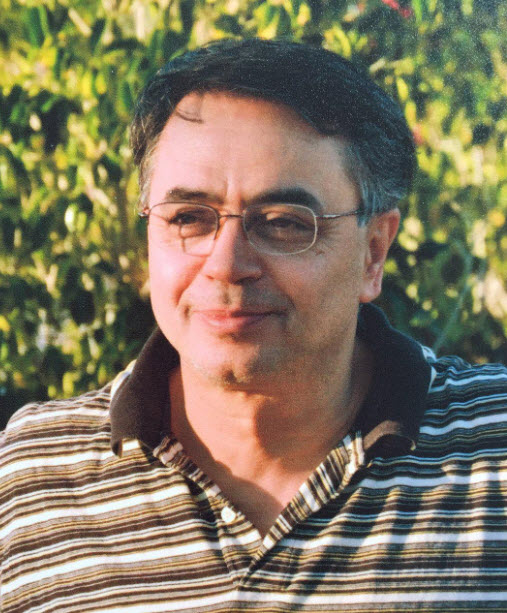}}]{Jawad A. Salehi}
(M'84-SM'07-FM'10) was born in
Kazemain, Iraq, on December 22, 1956. He received
the B.Sc. degree from the University of California,
Irvine, in 1979, and the M.Sc. and Ph.D. degrees from
the University of Southern California (USC), Los
Angeles, in 1980 and 1984, respectively, all in electrical
engineering. He is currently a Full Professor at
the Optical Networks Research Laboratory (ONRL),
Department of Electrical Engineering, Sharif University
of Technology (SUT), Tehran, Iran. From 1981
to 1984, he was a Full-Time Research Assistant at
the Communication Science Institute, USC. From 1984 to 1993, he was a
Member of Technical Staff of the Applied Research Area, Bell Communications
Research (Bellcore), Morristown, NJ. Prof. Salehi was an Associate
Editor for Optical CDMA of the IEEE Transactions on Communications,
since 2001 to 2012. He is among the 250 preeminent and most
influential researchers worldwide in the Institute for Scientific Information (ISI)
Highly Cited in the Computer-Science Category.
\end{IEEEbiography}
\end{document}